\documentclass[useAMS,usenatbib,usegraphicx]{emulateapj}
\usepackage{times}
\slugcomment{Astrophysical Journal, in press (Accepted 2013 Mar 04).}

\def\spitzer{{\em Spitzer}}
\def\wise{{\em WISE}}

\def\p{$\pm$}
\def\ltsim{\mathrel{\hbox{\rlap{\hbox{\lower4pt\hbox{$\sim$}}}\hbox{$<$}}}}
\def\gtsim{\mathrel{\hbox{\rlap{\hbox{\lower4pt\hbox{$\sim$}}}\hbox{$>$}}}}

\def\Msun{M$_{\odot}$}

\def\micron{$\mu$m}
\def\araa{ARA\&A}
\def\aap{A\&A}

\def\aaps{A\&AS}
\def\mnras{MNRAS}
\def\apj{ApJ}
\def\apjs{ApJS}
\def\aj{AJ}
\def\apjl{ApJL}
\def\pasp{PASP}
\def\pasj{PASJ}

\def\iraf{{\sc iraf}}
\def\sextractor{{\sc SExtractor}}

\def\ha{H$\alpha$}
\def\hb{H$\beta$}

\def\nai{Na~{\sc i}}

\def\lbol{$L_{\rm Bol}$}

\def\av{$A_{\rm V}$}

\pdfoutput=1
\shorttitle{Optical and mid-infrared evolution of SN~2009js}
\shortauthors{P. Gandhi et al.}

\begin{document}

\title{SN~2009js at the crossroads between normal and subluminous Type IIP supernovae: optical and mid-infrared evolution}

\author{P. Gandhi\altaffilmark{1,2}, M. Yamanaka\altaffilmark{3,4,5}, M. Tanaka\altaffilmark{6}, T. Nozawa\altaffilmark{7}, K.S. Kawabata\altaffilmark{5}, I. Saviane\altaffilmark{8}, K. Maeda\altaffilmark{7}, T.J. Moriya\altaffilmark{7,9,10}, T. Hattori\altaffilmark{11}, M. Sasada\altaffilmark{12}, R. Itoh\altaffilmark{3}
}
\altaffiltext{1}{Institute of Space and Astronautical Science, Japan Aerospace Exploration Agency, 3-1-1 Yoshinodai, chuo-ku, Sagamihara, Kanagawa 252-5210, Japan}
\altaffiltext{2}{Department of Physics, Durham University, South Road, Durham DH1 3LE, UK}
\altaffiltext{3}{Department of Physical Science, Hiroshima University, Kagamiyama 1-3-1, Higashi-Hiroshima 739-8526, Japan}
\altaffiltext{4}{Kwasan Observatory, Kyoto University, 17-1 Kitakazan-ohmine-cho, Yamashina-ku, Kyoto, 607-8471, Japan}
\altaffiltext{5}{Hiroshima Astrophysical Science Center, Hiroshima University, Higashi-Hiroshima, Hiroshima 739-8526, Japan}
\altaffiltext{6}{National Astronomical Observatory, Mitaka, Tokyo, Japan}
\altaffiltext{7}{Kavli Institute for the Physics and Mathematics of the Universe, University of Tokyo, Kashiwa, Japan}
\altaffiltext{8}{European Southern Observatory, Alonso de Cordova 3107, Santiago 19, Chile}
\altaffiltext{9}{Research Center for the Early Universe, Graduate School of Science, University of Tokyo, Hongo 7-3-1, Bunkyo, Tokyo 113-0033, Japan}
\altaffiltext{10}{Department of Astronomy, Graduate School of Science, University of Tokyo, Hongo 7-3-1, Bunkyo, Tokyo 113-0033, Japan}
\altaffiltext{11}{Subaru Telescope, National Astronomical Observatory of Japan, Hilo, Hawaii 96720, USA}
\altaffiltext{12}{Department of Astronomy, Graduate School of Science, Kyoto University, Kyoto 606-8502, Japan}

\label{firstpage}
\begin{abstract}
We present a study of SN~2009js in NGC 918. Multi-band Kanata optical photometry covering the first $\sim$120 days show the source to be a Type IIP SN. Reddening is dominated by that due to our Galaxy. One-year-post-explosion photometry with the NTT, and a Subaru optical spectrum 16 days post-discovery, both imply a good match with the well-studied subluminous SN~2005cs. The plateau phase luminosity of SN~2009js and its plateau duration are more similar to the intermediate luminosity IIP SN~2008in. Thus, SN~2009js shares characteristics with both subluminous and intermediate luminosity SNe. Its radioactive tail luminosity lies between SN~2005cs and SN~2008in, whereas its quasi-bolometric luminosity decline from peak to plateau (quantified by a newly-defined parameter $\Delta$log${\cal L}$ measuring adiabatic cooling following shock breakout) is much smaller than both the others. We estimate the ejected mass of $^{56}$Ni to be low ($\sim$0.007~\Msun). The SN~explosion energy appears to have been small, similar to SN~2005cs. SN~2009js is the first subluminous SN~IIP to be studied in the mid-infrared. It was serendipitously caught by \spitzer\ at very early times. In addition, it was detected by \wise\ 105 days later with a significant 4.6~\micron\ flux excess above the photosphere. The infrared excess luminosity relative to the photosphere is clearly smaller than that of SN~2004dj extensively studied in the mid-infrared. The excess may be tentatively assigned to heated dust with mass $\sim$3$\times$10$^{-5}$~\Msun, or to CO fundamental emission as a precursor to dust formation.
\end{abstract}

\section{Introduction}

Core-collapse supernovae (SNe) represent the deaths of massive stars through runaway implosions of cores, followed by explosive expulsion of their outer layers. They exhibit a diverse range of properties in terms of their spectral and photometric evolution \citep[e.g. ][]{filippenko97}. Observational follow-up within the past few years has also uncovered an incredibly broad range of emitted luminosities: from superluminous SNe \citep{galyam12} with luminosities greater than 7$\times$10$^{43}$ erg s$^{-1}$, to \lq subluminous\rq\ events \citep{pastorello04} that are factors of 10--100 times less powerful at peak. Understanding the origin of this diversity is an important goal of present research. 

The Type II class of SNe events display spectroscopic signatures of Hydrogen, and appear in at least three sub-classes. The IIP sources are named after a stage of flat (\lq Plateau\rq) light curves showing constant flux for several months. IIL sources, on the other hand, show a monotonic and linear decline of their post-explosion brightness without any plateau. Lastly, the IIn sub-class shows spectral features that are much narrower than in the other sub-classes, thought to be a result of strong interaction between SN ejecta and the circumstellar medium. Other sub-classes include the IIb (with weak hydrogen spectral features that soon disappear), and ambiguous events collected under the generic II-pec, or peculiar, sub-class.

The IIP population is important because it represents the most numerous sub-class, constituting about 75\% of all Type II SNe \citep{li11,smith11}. This sub-class itself encompasses a wide range of properties \citep{hamuy03}, and there is much debate regarding the progenitors of these events with present constraints favoring masses of between 8$\sim$16 \Msun\ \citep[e.g. ][ for a review]{smartt09}. The subluminous SNe may occupy the lower mass end of this range \citep{fraser11}, though comparatively few firm constraints exist. Their low luminosities have been attributed either to low mass progenitors, or to \lq fall back\rq\ events in which a massive progenitor gives rise to a black hole core and swallows a large fraction of the stellar material which would otherwise have escaped, thus limiting the observed luminosity \citep[e.g. ][]{zampieri98, moriya10}. Only about a dozen subluminous IIPs have been studied in any detail \citep[e.g. ][]{turatto98,pastorello04,spiro09,fraser11}, and understanding their nature will require observations of more sources to enlarge the present sample and also to bridge the \lq gap\rq\ between the subluminous and normal IIP population \citep[e.g. ][]{roy11}.

In this paper, we present optical and mid-infrared observations of SN~2009js in the galaxy NGC~918. As we show in this work, the source was a Type IIP event, which shares many characteristics with two other SNe: the well-studied subluminous SN~2005cs, and the intermediate luminosity SN~2008in. We report optical photometric evolution over the first year of SN~2009js, in addition to optical spectroscopy early in the plateau phase. From the multi-band light curves, we can constrain the explosion epoch, and measure the plateau length and source luminosity, thus allowing a computation of the line-of-sight reddening and explosion energy. Mid-infrared (MIR) photometric data with the \spitzer\ and \wise\ missions are also reported, and allow tentative constraints on the dust surrounding SN~2009js. As far as we are aware, this is the first mid-infrared study of a SN which shares characteristics of subluminous events. This work is thus a valuable addition to the low luminosity class of events. 

The discovery of SN~2009js was first reported by \citet{nakano09}, and also independently by \citet{silverman09} dating from 2009 Oct 11.44 and Oct 11.689. The SN was located at J2000 coordinates of RA=02h25m48.3s Dec=+18$^\circ$29$\arcmin$26$\arcsec$, about 35\farcs 5 W and 20\farcs 7 S of the nucleus of NGC 918. Optical spectra obtained the following day with the 3-m Shane reflector (+ Kast) at Lick Observatory indicated the object to be very similar to the Type IIP SN~2005cs about two days post-maximum brightness. An \ha\ absorption blueshift of 7200 km s$^{-1}$ intrinsic to the SN was measured in this spectrum. 

We assume a distance to the host galaxy NGC~918 of 21.7\p 1.8 Mpc ($z$=0.005; e.g. \citealt{martinezgarcia09} and references therein). This galaxy is of morphological class Sc\footnote{http://leda.univ-lyon1.fr} oriented at a position angle of 158$^\circ$. It has hosted two SNe in the past three years: SN~2011ek (a Type Ia reported by \citealt{nakano11}), and SN~2009js which is the subject of this paper.


\section{Observations}

\subsection{Optical imaging and spectroscopy}

Three sets of optical imaging data are used in our work: 

\begin{enumerate}

\item Early-time photometry, including pre-discovery limits and discovery data \citep{nakano09, silverman09}. These were obtained in filterless imaging mode. Throughout this paper, we assume a discovery epoch of 2009 Oct 11.44 \citep{nakano09, silverman09}. 

\item Plateau-phase $BVR_{\rm C}I_{\rm C}$ follow-up of the field of SN~2009js carried out with the HOWPol instrument (Hiroshima One-shot Wide-field Polarimeter; \citealt{howpol}) at the Kanata telescope, starting on 2009 Oct 14.8 at an epoch of +3.4 days (epochs are quoted post-discovery throughout) and continuing for about 4 months. 

\item A single late-time epoch in the $VRI$ filters is publicly available from observations made at the ESO Faint Object Spectrograph and Camera (EFOSC; \citealt{efosc}) mounted on the New Technology Telescope (NTT). These data date from 2010 Oct 6, just under one year post-explosion.

\end{enumerate}

All observation details (including logs and analysis) may be found in the Appendix. The photometric measurements are listed in Table~\ref{tab:kanata}. 

In addition to photometry, an optical spectrum of SN 2009js was obtained by some of us on UT 2009-10-27 (MJD = 55131.6), about 16 days post-discovery, with the Faint Object Camera and Spectrograph (FOCAS; \citealt{focas}) mounted on the Subaru telescope. This provides important information on the SN type and ejecta kinematics. Observational details may again be found in the Appendix.

\subsection{Mid-infrared imaging}

Although the source had been monitored in the optical since 2009, we identified it as being of particular interest in 2011 when we noticed its detection in the mid-infrared serendipitously by two missions. MIR detections of SNe (especially of the low luminosity class that SN~2009js belongs to) are relatively scarce. 

\spitzer\ observed the host galaxy NGC~918 for unrelated science\footnote{Cycle 9 PROGID 61060, P.I. K. Sheth.} on two occasions: about one month pre-explosion and also two days post-discovery. SN~2009js shows a bright detection in the latter set of images in both observed bands centered on 3.6 and 4.5 \micron\ (hereafter, channels IRAC1 and IRAC2, respectively) of the InfraRed Array Camera (IRAC; \citealt{irac}).

On longer timescales, the {\em Wide-field Infrared Survey Explorer} (\wise) satellite \citep[][]{wise} detected the source in its W1 and W2 bands centered on wavelengths of $\approx$3.4 and 4.6~\micron\ respectively, at an epoch of about +107 days. These data are publicly-available as part of the mission all-sky survey release. Later non-detections at epochs of about +295 and +470 days are also available. 

\begin{figure*}
  \begin{center}
    \includegraphics[width=12.5cm,angle=0]{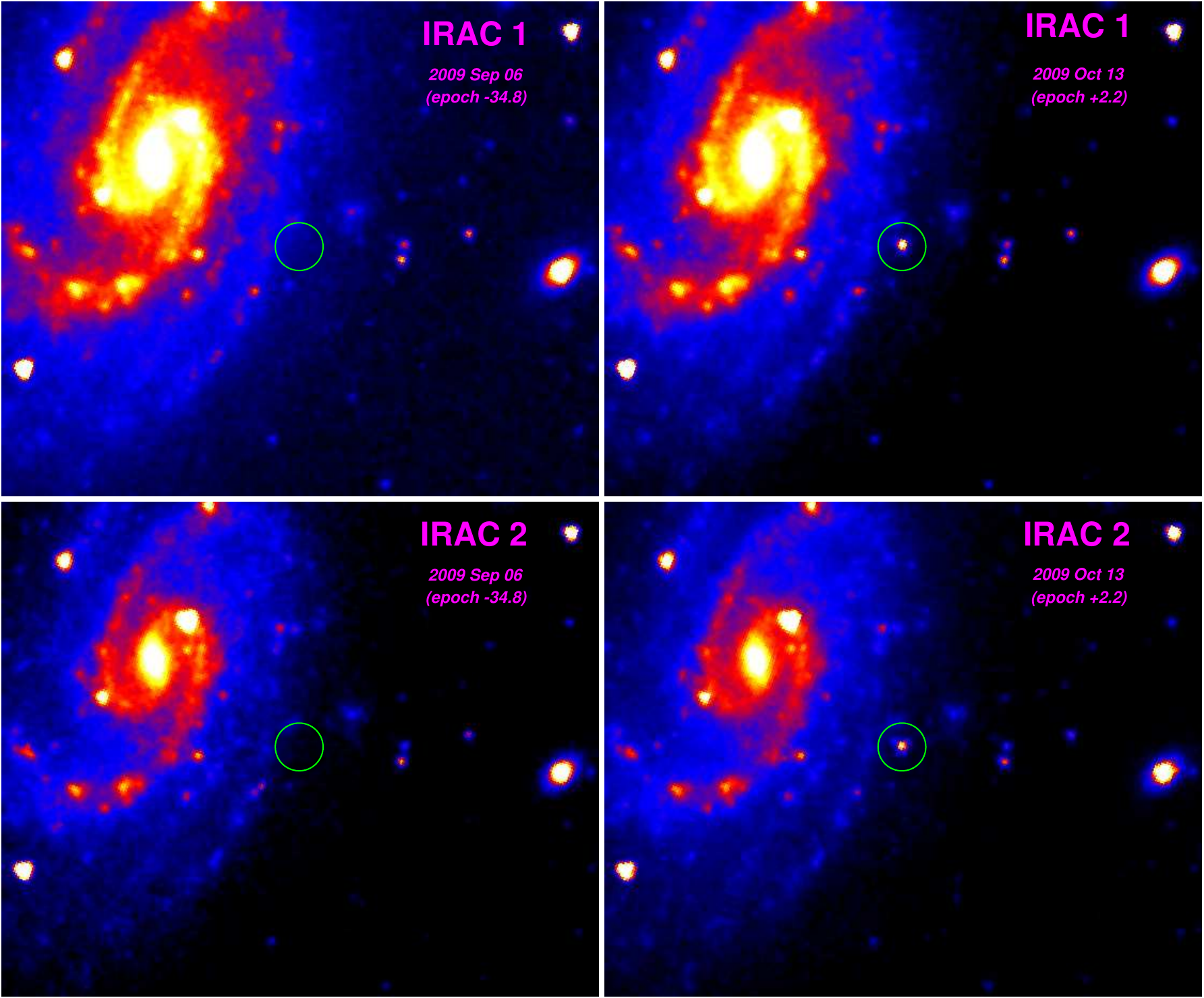}
    \caption{\spitzer\ IRAC1 and IRAC2 pre-explosion images from 2009 Sep 06 at an epoch of 34.8 days before discovery (left column), and 2.2 days post-discovery on 2009 Oct 13 (right column). The position of SN~2009js is marked by the green circle. The size of each image is about 2.4\arcmin$\times$2\arcmin, with North being up and East to the left. The SN is clearly visible just two days after explosion in the column on the right in both bands.
    \label{fig:spitzer}}
  \end{center}
\end{figure*}

\begin{figure*}
  \begin{center}
    \includegraphics[width=12.5cm,angle=0]{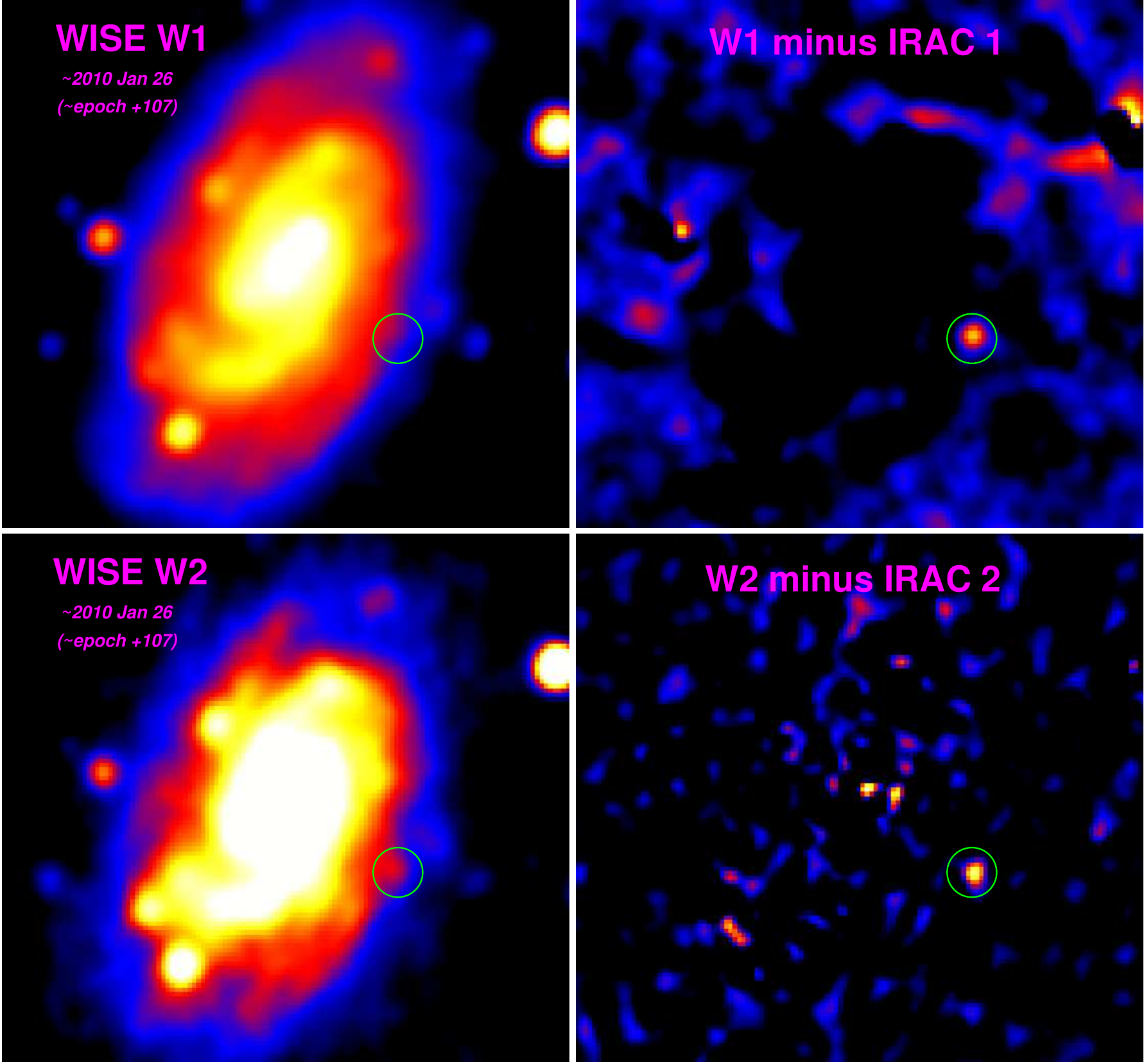}\hspace{5cm}
    \includegraphics[width=4.5cm,angle=0]{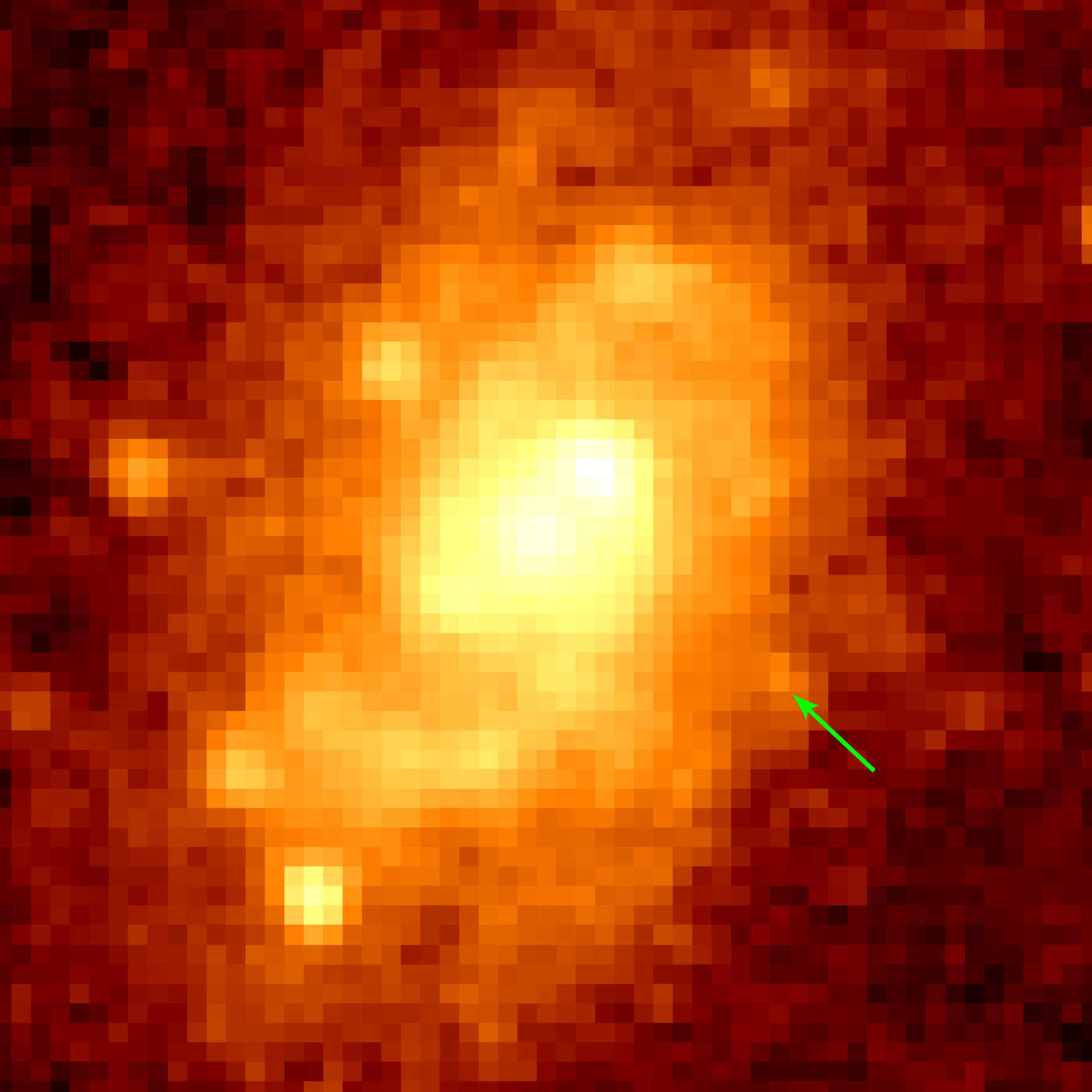}
    \includegraphics[width=4.5cm,angle=0]{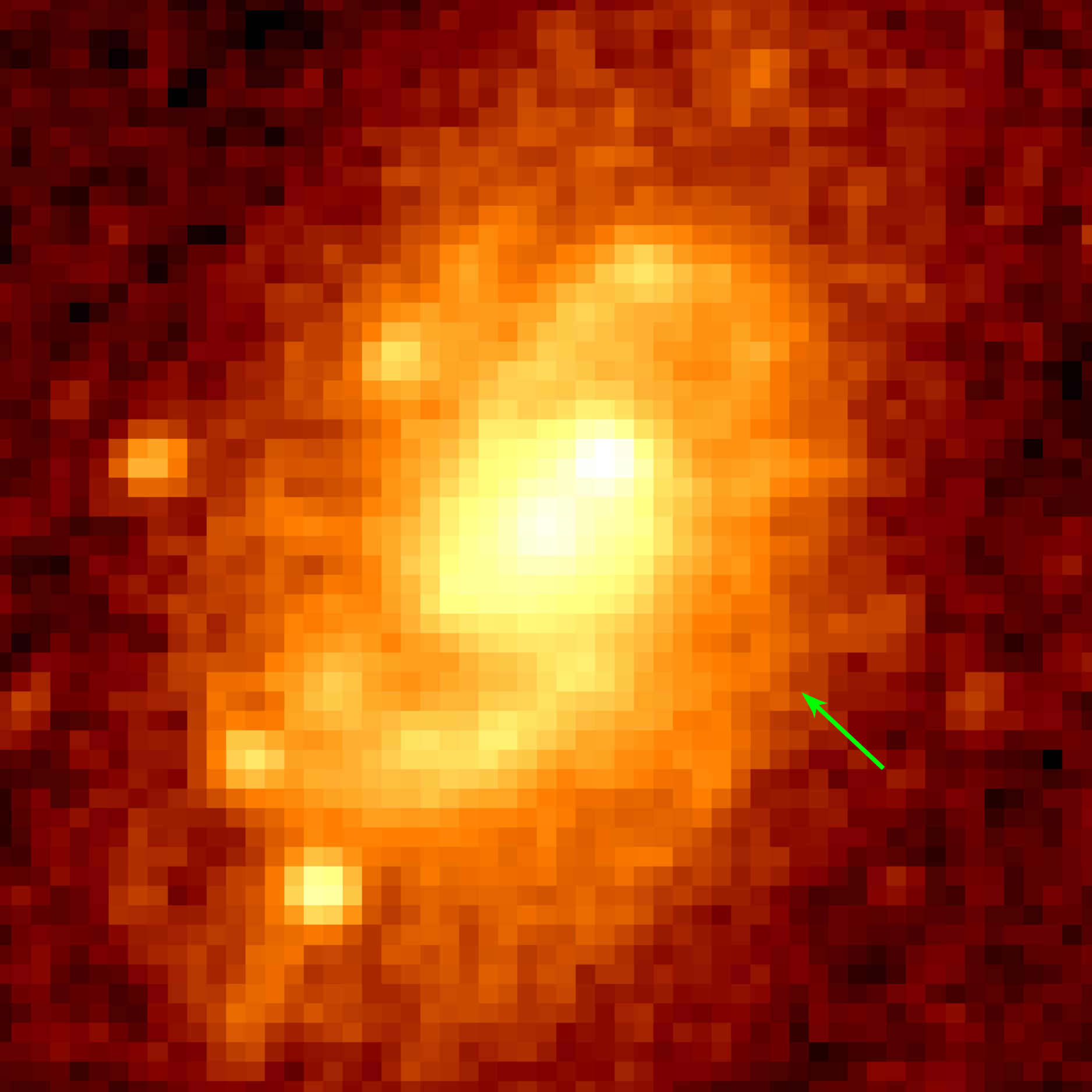}
    \includegraphics[width=4.5cm,angle=0]{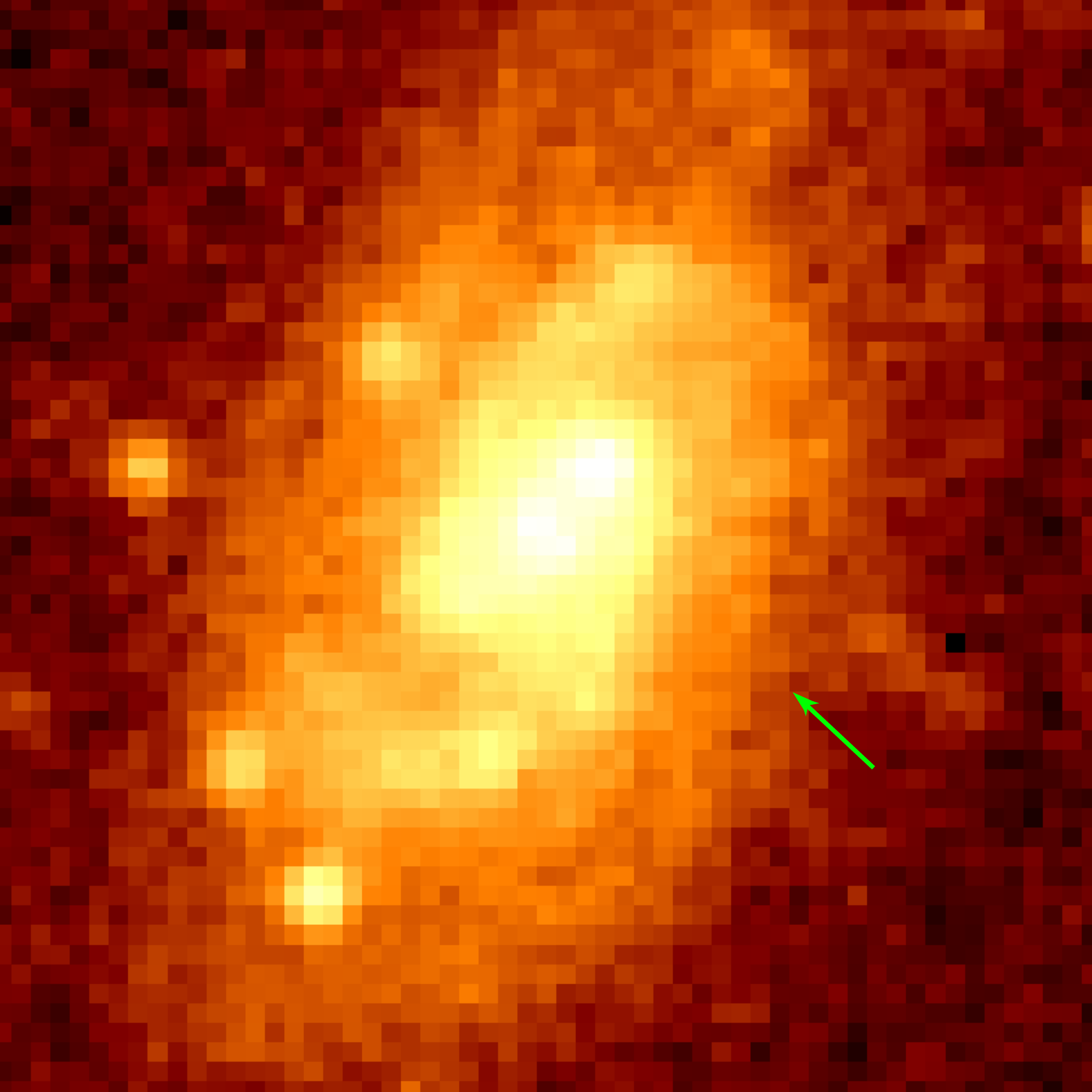}
    \caption{(Top and Middle) \wise\ 3.4 and 3.6~\micron\ (W1 and W2) Atlas L3 images of NGC 918 taken around 2010 Jan 26 (epoch $\sim$107 days post-discovery) are shown in the left column. In the right column, \spitzer\ pre-explosion IRAC1 and IRAC2 images have been subtracted, after scaling and PSF matching. The SN is clearly visible. (Bottom) Coadded images produced from L1 data of each of the three \wise\ epochs (see Table~\ref{tab:obslog}) shown in order from left to right, with the SN position marked by a green arrow. North is up and East left, and images are approximately 2\farcm 5 on a side.
    \label{fig:wise}}
  \end{center}
\end{figure*}

The MIR observation log may be found in the Appendix (Table~\ref{tab:obslog}) along with a full description of the data analysis and photometry procedures. MIR flux measurements are listed in Table~\ref{tab:irphot}. \spitzer\ and \wise\ images are shown in Figs.~\ref{fig:spitzer} and \ref{fig:wise}, respectively.

\section{Results}
\label{sec:results}

\subsection{Reddening due to the Galaxy}
\label{sec:galreddening}

The optical data were dereddened assuming a standard interstellar extinction law with $R_{\rm V}$=3.1 \citep{fitzpatrick99}. The Galactic extinction along this line-of-sight is determined to be \av=0.95 from the work of \citet{schlafly11}. A 10\% uncertainty is assumed on this value. 

In the mid-infrared, several more recent determinations of the dereddening law have been published, and we used an average extinction law computed from these, resulting in $A_{3.4\ \mu{\rm m}}$/\av=0.07 and $A_{4.6\ \mu{\rm m}}$/\av=0.05 (See \S~2.2.3 of \citealt{g11_wise} for details and references).

\subsection{Reddening local to the host galaxy}
\label{sec:reddening}

Extinction local to the source was estimated by using the relationship between host galaxy reddening and plateau $V$--$I$ color identified by \citet{olivares10}. The $V$--$I$ color is that measured at --30 days from the mid-point of the transition phase ($t_{\rm PT}$). The optical light curves and $t_{\rm PT}$ measurement is described in the following section. $t_{\rm PT}$--30 corresponds to day 80, at which time the $V$--$I$ color of SN~2009js (corrected for Galactic reddening) is 0.73\p0.13. Using Eq. 7 of \citet{olivares10} returns \av(host)=0.18\p0.38.

\subsection{Optical light curves}
\label{sec:optlc}

\begin{figure*}
  \begin{center}
  \includegraphics[angle=0,width=14.5cm]{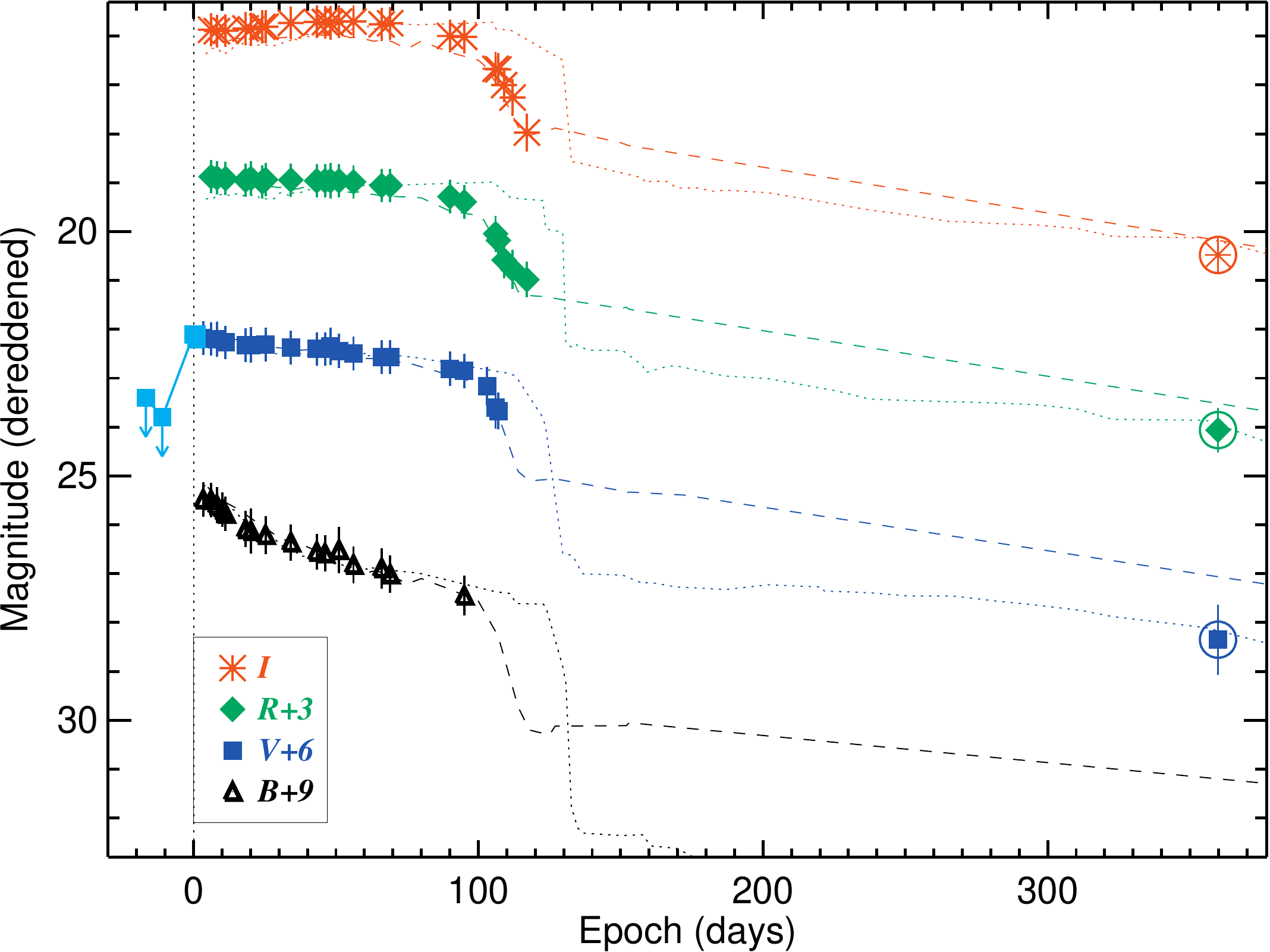}
  \caption{Reddening-corrected $BVRI$ light curves of SN~2009js compared to those of SN~2005cs (thin dotted curves) and SN~2008in (dashed) covering both the Kanata and NTT monitoring data to an epoch of about 360 days. Both Galactic and local reddening corrections have been applied for all objects. The cyan upper-limit arrows and detections on the left denote the pre-explosion limits and early time photometry from the literature. This is unfiltered photometry, closely matched to the $V$ band (see Appendix for details). The circled symbols on the right are the NTT $VRI$ magnitudes. Light curves for SN~2005cs and SN~2008in have been scaled in order to approximately match during plateau, for comparison. 
    \label{fig:fulllc}}
  \end{center}
\end{figure*}

The change in brightness between the early-time pre-discovery and the discovery epoch photometry implies a brightening by at least 1.7 mags in less than 11 days (see Table~\ref{tab:kanata}). Fig.~\ref{fig:fulllc} shows the long-term source optical light curves. The early-time photometric detections from published filterless photometry match our first Kanata $V$ band measurements closely. The long steady light curves in several (redder) Kanata filters are characteristic of Type IIP supernovae. 
The late-time NTT magnitudes probe the nebular phase and show a dimming with respect to the peak brightness measured by Kanata of about 6.2, 5.2 and 4.8 mags in $V$, $R$ and $I$, respectively.

In order to characterize the light curve evolutionary phases, we fitted the latter part of the $I$ band light curve (beyond an epoch of +60 days) using a two component Fermi-Dirac function plus linear function following the procedure of \citet[][ see their Eqs. 1 and 2]{olivares10}. The $I$ band was chosen because the comparison with the light curves of SN~2005cs in Fig.~\ref{fig:fulllc} shows that this band probably covers the bulk of the transition phase between the plateau and tail (this will be furthered discussed in \S~\ref{sec:nebular}). Although our Kanata data do not cover the beginning of the tail, the NTT photometry allows an approximate (i.e. two-point) estimation of the evolution during this phase, under the assumption that the Kanata data cover the bulk of the transition. The best-fit parameters for the transition phase are found to be 

\begin{eqnarray} 
  a_0 & = & 2.24\pm 1.04\ {\rm mag} \\
  t_{\rm PT} & = & 111.3\pm 3.6\ {\rm days}\\
  w_0 & = & 4.5\pm 2.4\ {\rm days}\\
  p_0 & = & 0.009\pm 0.004\ {\rm mag\ day}^{-1}\\
  m_0 & = & 18.46\pm 0.93\ {\rm mag}\\
  \chi^2/{\rm dof} & = & 1.7/5
\end{eqnarray}

where $a_0$ represents the plateau brightness above the linear phase (in mag), $t_{\rm PT}$ denotes the mid-point of the transition phase in days (from discovery) and $w_0$ is the width of this phase (in days). The slope and intercept of the linear radioactive nebular decline are $p_0$ and $m_0$, respectively. The goodness of fit is quantified by $\chi^2/{\rm dof}$, with dof being the degrees of freedom.

\subsection{Mid-Infrared photometry}
\label{sec:results_irphot}

The MIR detection of SN~2009js was first noticed by us during an examination-by-eye of the \wise\ all-sky preliminary data release fields (first-pass processing prior to the all-sky release) of supernovae to have occurred during, or before the beginning of, the \wise\ survey. Thereafter, we checked and found the source to be detected with \spitzer\ as well. 

The \spitzer\ detection is very significant at both the observed wavelengths of 3.6 and 4.5 \micron\ two days post-discovery. As compared to the pre-explosion images obtained about 37 days prior to this, the flux at the position of the SN increased by factors of at least 25 and 11 in the two channels, respectively (Table~\ref{tab:irphot}). This is undoubtedly a strong lower limit to the change in flux of the SN itself. 

The \wise\ data show the source to be clearly present in the Atlas images taken about 100 days post-discovery. Table~\ref{tab:irphot} implies a decline in flux by factors of 1.8 and 1.6 (or magnitudes of 0.6 and 0.5) in W1 and W2 over this period, as compared to \spitzer\ IRAC1 and IRAC2, respectively. The \wise\ detection is highlighted in Fig.~\ref{fig:wise} by subtracting smoothed and resampled IRAC1 and IRAC2 pre-explosion images from the Atlas W1 and W2 images, respectively. Proximity to the host galaxy results in non-uniform underlying residuals. The combination on an uneven background with the large \wise\ PSF may explain why the source was not picked up by the automated pipeline. 

The Atlas image presented in Fig.~\ref{fig:wise} is made from a combination of all images from the first epoch (2010 Jan) of \wise\ imaging and some images from the second epoch (2010 Aug). Comparison with our coadded images produced from the L1b frames (see description in \S~\ref{sec:wise} and lower panels shown in Fig.~\ref{fig:wise}), we find that the signal in the Atlas images is completely dominated by first epoch (2010 Jan). 

Subsequent observations by \wise\ (in 2010 Aug and 2011 Jan) do not show any appreciable detections. Detection upper limits at epochs of $\sim$300 and 470 days post-discovery in Table~\ref{tab:irphot} show that the SN declined by a factor of at least 1.4 in flux (and probably more) in both bands on timescales of $\sim$6 months or less.

\subsection{Optical spectrum}
\label{sec:optspec}

\begin{figure*}
  \begin{center}
    \includegraphics[angle=0,width=12.5cm]{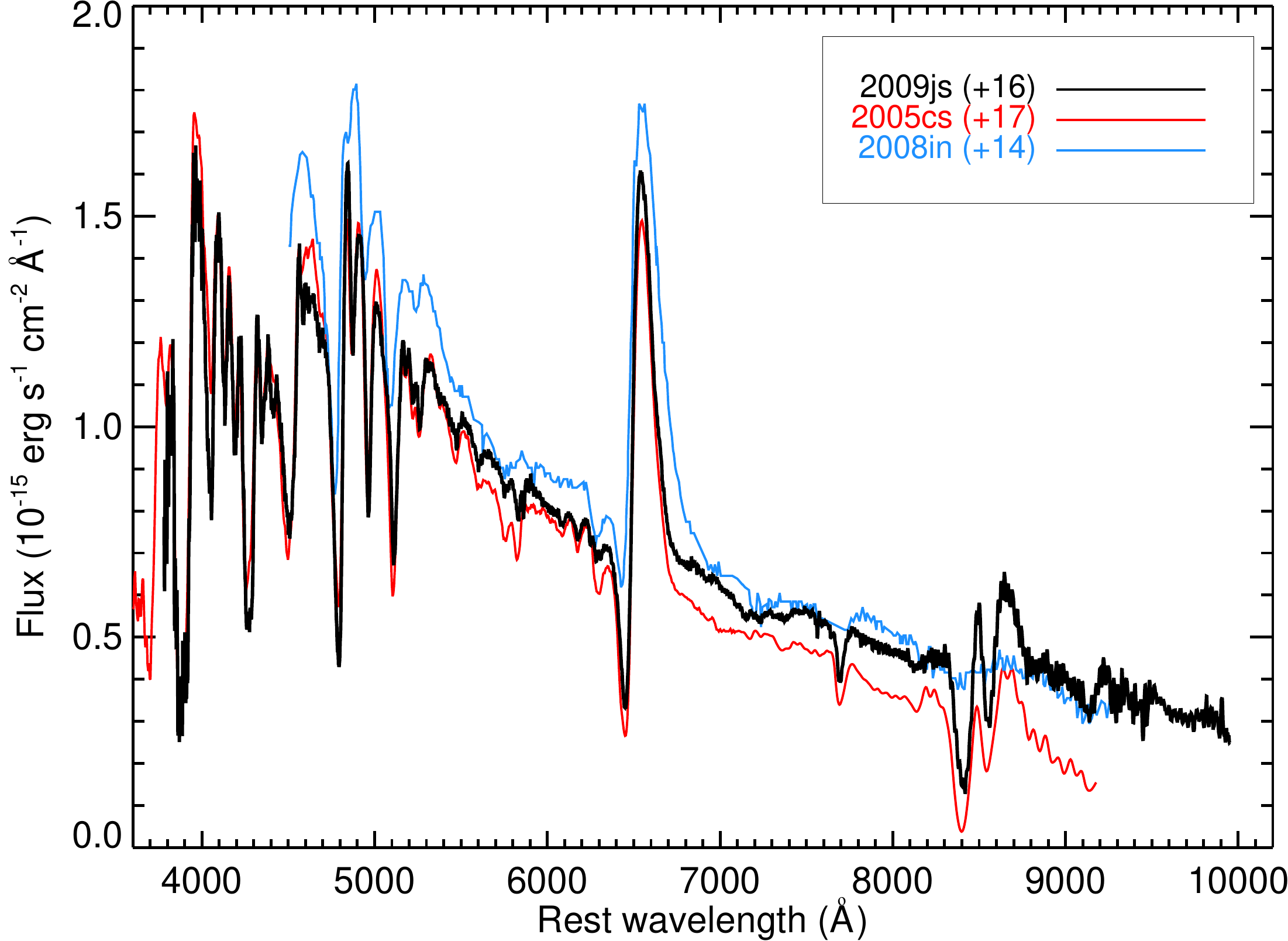}
    \caption{Subaru/FOCAS spectrum of SN 2009js at 16 days post-discovery (black) compared to the Mt. Ekar 1.82m AFOSC spectrum of SN~2005cs 17 days post-discovery (red curve observed on 2005-07-14; from \citealt{pastorello09}), and with SN~2008in 14 days post-explosion (blue curve observed on 2009-01-07; from \citealt{roy11}). The comparison spectra of SN~2005cs and SN~2008in were normalized to the continuum of SN~2009js around 7000~\AA, and then offset slightly for better legibility. The match between SN~2009js and SN~2005cs is excellent, except that the continuum slope of SN~2005cs is slightly bluer as a result of its hotter temperature. SN~2008in was further scaled in order to match the peak strength of the \ha\ emission line. SN~2008in shows a significant difference with respect to the other two objects in terms of having a much shallower blueshifted \ha\ absorption feature. 
    \label{fig:spec}}
  \end{center}
\end{figure*}

The Subaru spectrum of SN~2009js is shown in Fig.~\ref{fig:spec}. The dereddened data have a blue spectral shape over the optical regime as expected from the hot temperature $T$$\approx$7000 K around 16 days post-discovery. A P Cygni profile of the absorbed blue wing for a strong emission line is most clearly visible for \ha, but also for \hb. Many other significant absorption features can be discerned over the full wavelength range. 

The \nai\ D doublet near 5891~\AA\ is a tracer of line-of-sight extinction, and our spectrum shows two \nai\ absorption features: one in the restframe of the host galaxy, and one consistent with extinction arising in the Milky Way. The equivalent widths of the features are $\approx$1.1~\AA\ and 3.6~\AA, respectively. The Milky Way feature is the stronger of the two, which is consistent with the reddening trend determined in \S \ref{sec:galreddening} and \ref{sec:reddening}. 

In order to classify the SN spectrum, we used the {\sc snid} code \citep{snid} in order to compare this source to previous events. The returned closest match is to the Type IIP SN~2005cs \citep{pastorello09}. The optical spectrum of this source at a phase of $\approx$17 days is overplotted in Fig.~\ref{fig:spec} and shows an excellent match to SN~2009js. Measuring the centroid wavelength of the trough of the \ha\ P Cygni absorption component in the restframe spectrum yields a blueshift of 5216\p30 km s$^{-1}$ for SN~2009js.

\section{Discussion}

\subsection{Explosion epoch}
\label{sec:explosionepoch}

The exact explosion epoch of SN~2009js is unknown. Early-time discovery epoch photometry aligns well with our subsequent Kanata monitoring meaning that we do not probe the initial flux rise. Using the non-detection at an epoch of --11 days (Table~\ref{tab:kanata}) the explosion can be constrained to have occurred at MJD 55109.94 (2009 Oct 05.94) to within an uncertainty of 5.5 days. 

\subsection{Evolution of temperature and bolometric luminosity}
\label{sec:evolution}

\begin{figure}
  \begin{center}
  \includegraphics[angle=0,width=8.5cm]{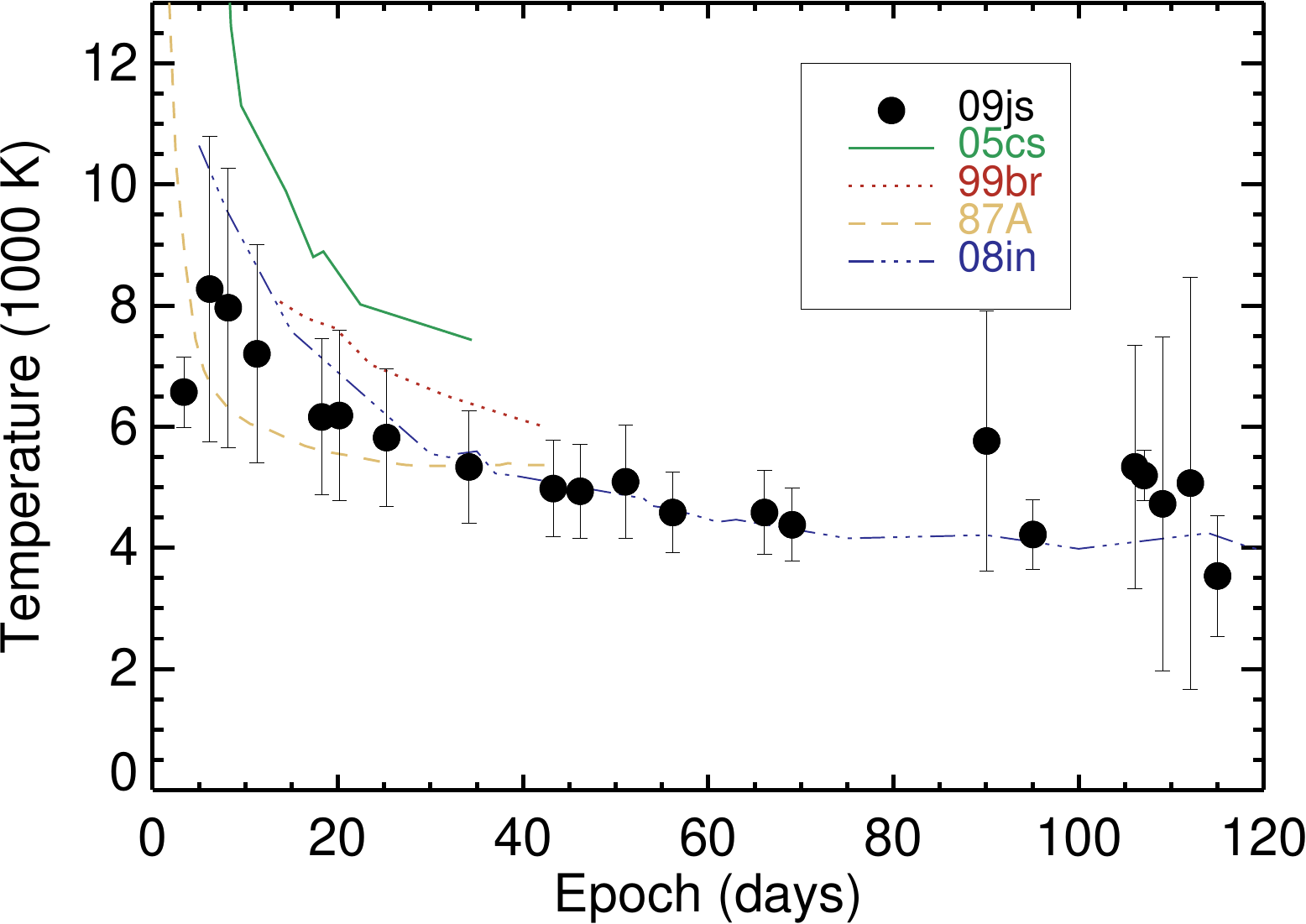}
    \caption{Evolution in blackbody temperature after correction for local and Galactic reddening. 
    \label{fig:temperature}}
  \end{center}
\end{figure}

\begin{figure*}
  \begin{center}
  \includegraphics[angle=0,width=12.5cm]{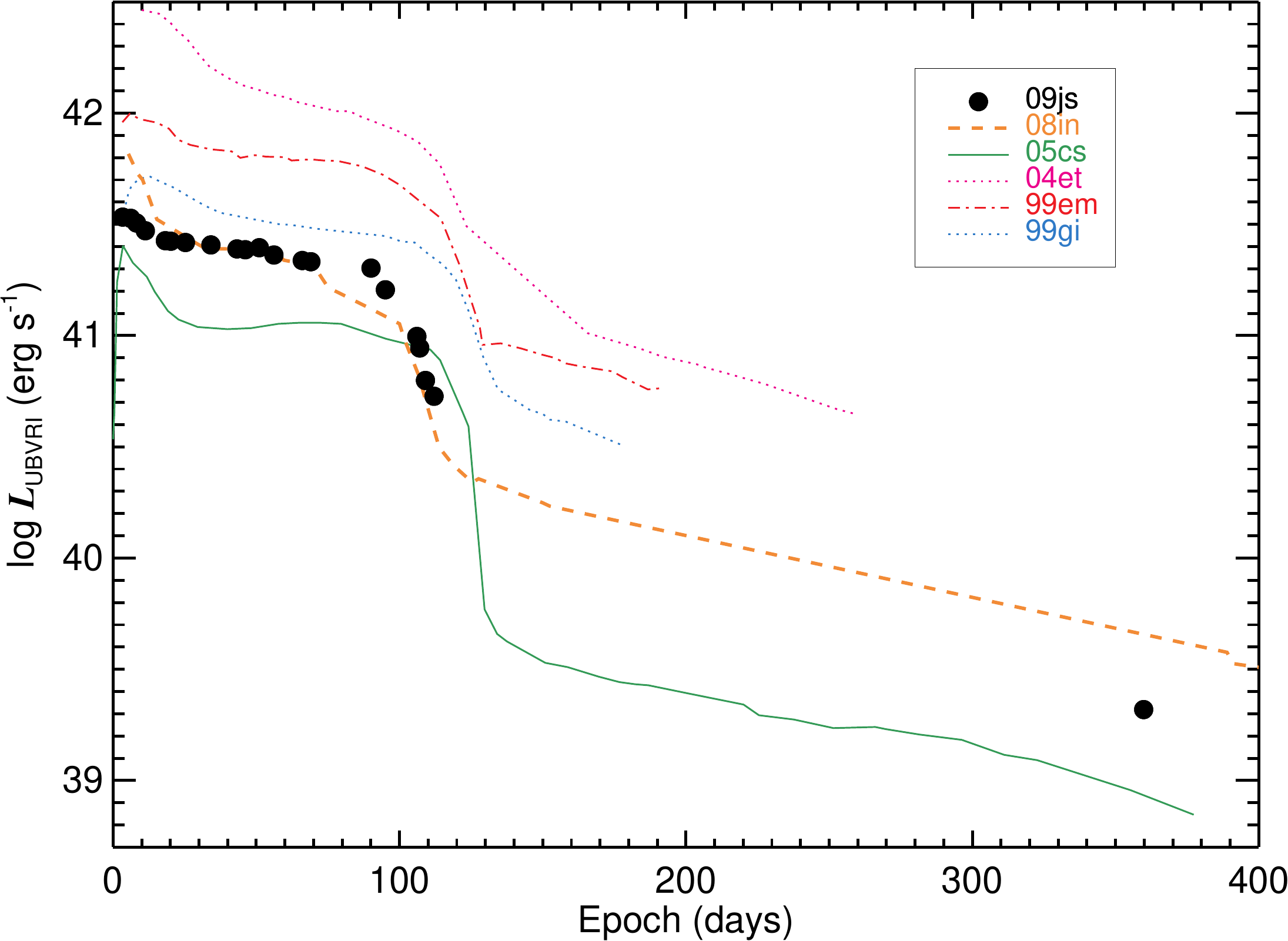}
    \caption{Evolution in quasi-bolometric power (integrated under the Planck function between 3500\AA\ and 1~\micron) after correction for local and Galactic reddening. Data for comparison SNe are taken from Fig.~9 of \citet{pastorello09} except for SN~2008in, for which $L_{\rm UBVRI}$ is computed from the photometric data presented by \citet{roy11} in a manner identical to that for SN~2009js. 
    \label{fig:evolution}}
  \end{center}
\end{figure*}

\begin{figure*}
  \begin{center}
    \includegraphics[width=4.25cm,angle=0]{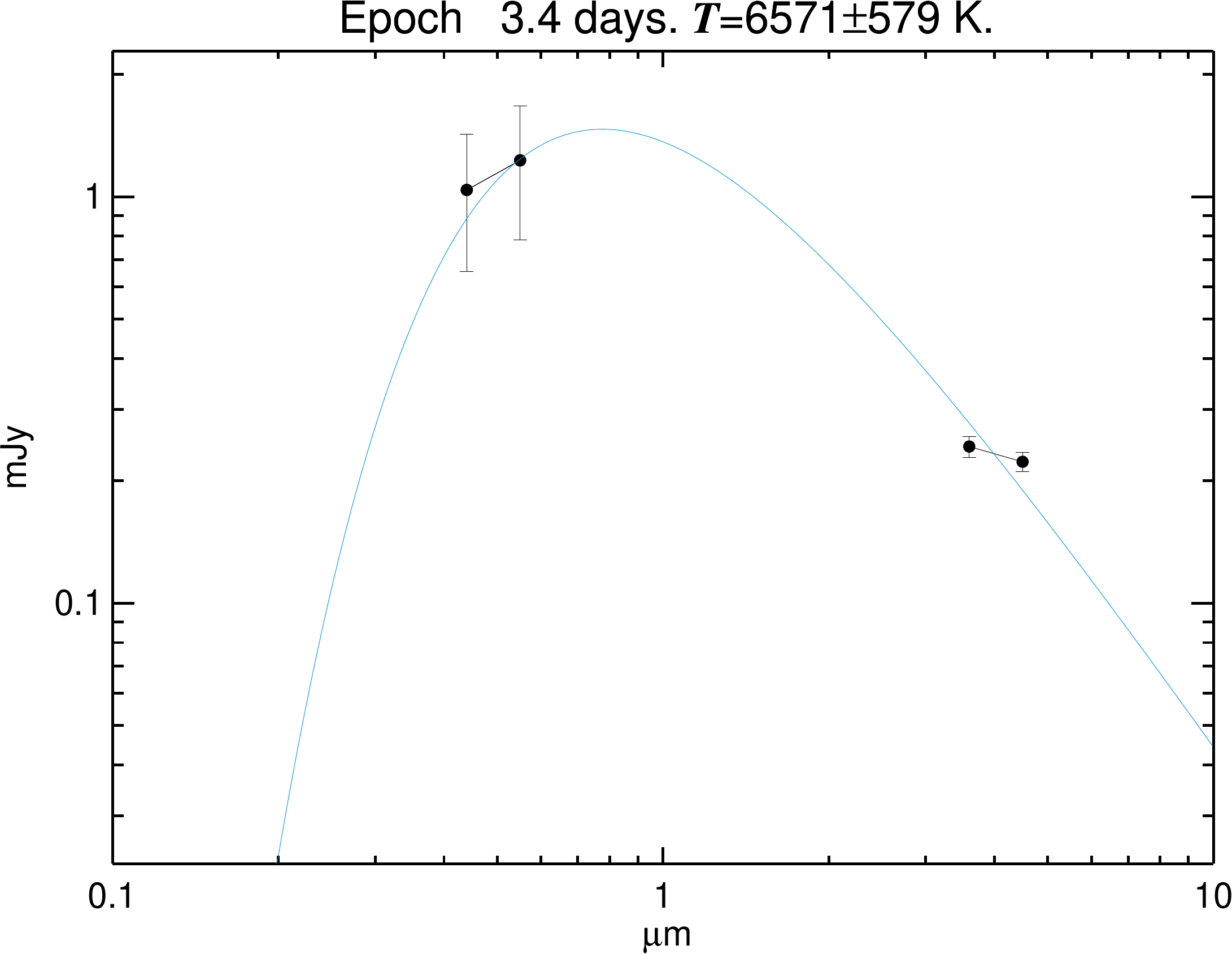}
    \includegraphics[width=4.25cm,angle=0]{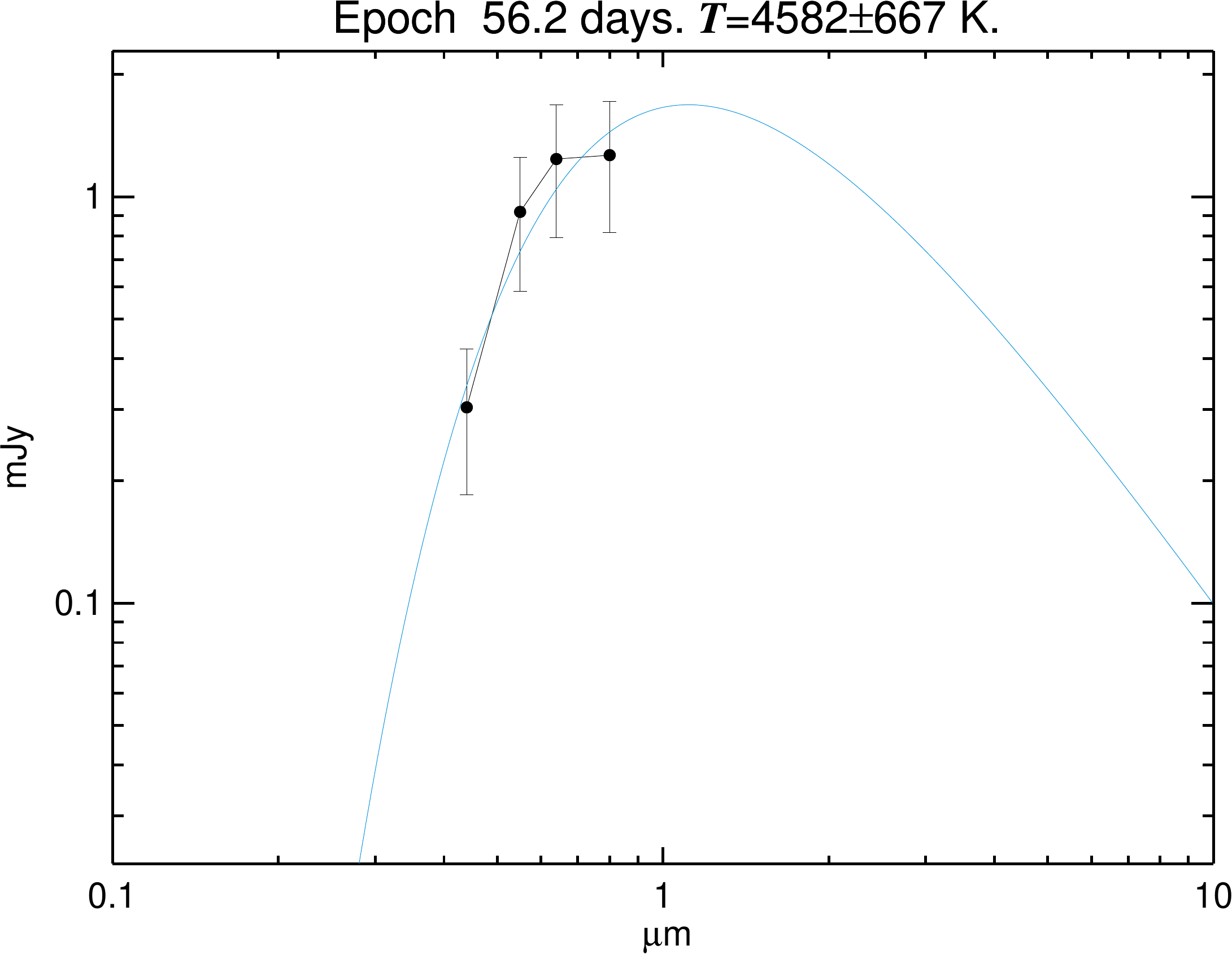}
    \includegraphics[width=4.25cm,angle=0]{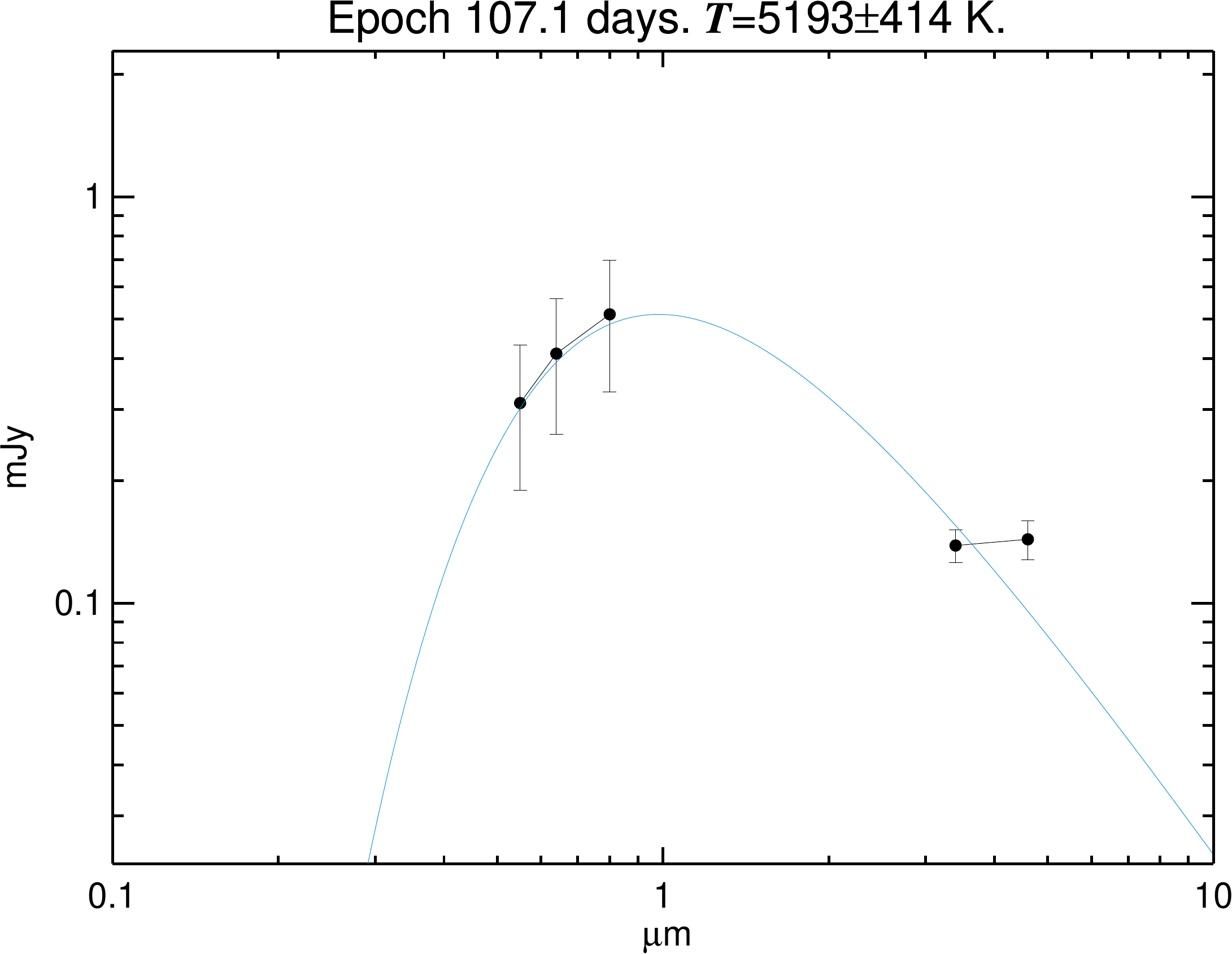}
\caption{Blackbody fits to multi-wavelength epoch SEDs, corrected for local and Galactic reddening. Three epochs close to the beginning (epoch +3.4), middle (epoch +56.2) and end (epoch +107.1) of Kanata monitoring are shown for illustration. The first and last of these include MIR detections. 
    \label{fig:epochs}}
  \end{center}
\end{figure*}

Multi-band photometry is available for multiple epochs of Kanata monitoring, and for one late epoch at the NTT (Table~\ref{tab:kanata}). Each epoch was corrected for reddening due to the Galaxy and that local to the source, and then fitted with single Planck functions under the assumption that the early evolution is dominated by photospheric emission. Of the 25 nights of optical Kanata photometry, epochs +3.4 and +107.1 coincide closely with the MIR detections of SN~2009js with \spitzer\ and \wise, respectively. In these two cases, the optical and MIR data were fitted simultaneously. Best fit parameters and uncertainties were measured using the {\sc mpfit} package \citep{mpfit} in {\tt idl}.\footnote{http://www.exelisvis.com/ProductsServices/IDL.aspx} This allows measurement of photospheric temperatures ($T$) and (quasi)bolometric luminosities ($L_{\rm UBVRI}$), both of which are plotted as functions of time in Figs.~\ref{fig:temperature} and \ref{fig:evolution} respectively. Three representative epoch fits are shown in Fig.~\ref{fig:epochs}. 

There is an apparent fast rise in $T$ between the first two Kanata temperature measurements (a timespan of three days), which is intriguing given that none of the light curves in Fig.~\ref{fig:fulllc} show any dramatic increase at this phase. Some caution is necessary in interpretation of this rise, given that it is only a single measurement of temperature change and the uncertainties are somewhat large. The subsequent evolution shows a decline from a peak of $\sim$8500 to $\ltsim$5000 K over a timespan of about 120 days. 
Data in all optical filters were not available for each epoch, resulting in large uncertainties when only a few filters were observed. 
But the systematic trend of temperature evolution is clear irrespective of these variations. 

Fig.~\ref{fig:evolution} shows the change in $L_{\rm UBVRI}$, the optical quasi-bolometric luminosity. Although we do not have $U$ band monitoring, $L_{\rm UBVRI}$ can be measured as the integration under the fitted Planck curve between wavelengths of 0.35--1~\micron. This parametrization of the optical bolometric power has been chosen to facilitate direct comparisons with other measurements in the literature, which will be discussed shortly. The peak luminosity is log $L_{\rm UBVRI}$ (erg s$^{-1}$)$\approx$41.6. This quasi-bolometric light curve also shows four distinct phases: 

\begin{enumerate}
\item an initial fast fade within the first $\sim$20 days;
\item a period of approximately constant output for the next $\sim$70 days;
\item then a sharp decline in the transition between the plateau and nebular phases to the last probed Kanata datapoint;
\item and finally, an apparently gentler decline between the Kanata and NTT epochs. 
\end{enumerate}

The mean change in $L_{\rm UBVRI}$ during these four periods are approximately: 1) --0.011; 2) --0.001; 3) --0.024 and 4) --0.006, as measured in log luminosity (dex) per day. 

The {\em total} bolometric power (\lbol) may be computed by integrating under the full Planck curve, including contributions from the far-ultraviolet and infrared regimes. This is a factor of 1.50 higher than $L_{\rm UBVRI}$, on average. Integrating \lbol\ over all epochs returns a total radiative energy of about 5.2$\times$10$^{48}$ erg.

\subsection{SN~2009js: comparisons with SN~2005cs and SN~2008in}
\label{sec:2005cslike}

The optical light curves discussed above (especially in the $R$ and $I$ bands, as well as in $L_{\rm UBVRI}$) show the presence of a clear plateau phase. This confirms the Type IIP nature of the source, initial suggestions of which were based upon very-early-time spectroscopy \citep{silverman09}. The long-term $VRI$ light curves shown in Fig.~\ref{fig:fulllc} suggest that the change in brightness from mid-plateau to late nebular phase is overall similar to that of the subluminous SN~2005cs.

Based upon early-time optical spectroscopy, \citet{silverman09} measured an \ha\ absorption redshift of about 7200 km s$^{-1}$ at two days after maximum brightness, and suggested a match to SN~2005cs. Peak brightness corresponds to a phase of 2 days according to \citet{pastorello09}. Our measurement of the blueshift in the Subaru spectrum obtained at a phase of about 16 days post-discovery is 2000 km s$^{-1}$ lower (\S~\ref{sec:optspec}), as expected from the gradual lowering with time of the outer photospheric layer that is optically-thick. This Subaru spectrum also displays remarkable affinity to that of SN~2005cs at 17 days (Fig.~\ref{fig:spec}). Although there is some uncertainty in determination of the explosion epoch of SN~2009js, these matches provide evidence (in addition to that discussed in \S~\ref{sec:explosionepoch}) that SN~2009js was discovered close in time of explosion (i.e. within a few days). All these lines of evidence suggest that SN~2009js was a 2005cs--like event. 

Both sources show peak optical bolometric luminosities of well below 10$^{42}$ erg s$^{-1}$, above which \lq normal\rq\ Type II SNe generally lie. Furthermore, both sources have absolute $V$--band magnitudes much lower (intrinsically fainter) than $M_V$=--16. This appears to support a subluminous classification for SN~2009js also. But Fig.~\ref{fig:evolution} shows that the plateau phase luminosity of SN~2009js is instead much more closely matched to that of another source, SN~2008in. This was also a Type IIP event, studied in detail by \citet{roy11}. It has been classified as an intermediate luminosity source, bridging the \lq gap\rq\ between the subluminous and normal populations. We computed $L_{\rm UBVRI}$(2008in) by using the multi-band photometry and extinction determination presented by \citet{roy11}, by using Planck-function fitting exactly as for SN~2009js. Almost identical results are obtained if we use the quasi-bolometric lightcurves (in terms of $UVOIR$) presented in Fig.~12 of \citet{roy11} and use a mean conversion factor to the $UBVRI$ wavelength range over the lightcurve. In addition, Fig.~\ref{fig:vabs} shows single-band absolute magnitude light curves, avoiding any uncertainties related to bolometric corrections, and again suggests a reasonable match between SN~2008in and SN~2009js during the plateau phase. On the other hand, the main difference with respect to SN~2008in is in terms of the spectra: Fig.~\ref{fig:spec} shows that significantly broader \ha\ in emission for SN~2008in and a blueshifted absorption feature which is weaker by a factor of about 2 (in terms of equivalent width) as compared to the other two sources. \citet{roy11} have also studied the evolution of the spectral velocities for SN~2008in and found significantly larger velocities than for SN~2005cs over the first 100 days.

Thus, it appears that SN~2009js shares mixed characteristics of both \lq subluminous\rq\ and \lq intermediate\rq\ luminosity events. We discuss other similarities and differences between these three sources in the following sections. Ultimately, however, the exact nomenclature of the assigned class is not crucial. Rather, the main question is whether the underlying progenitor and explosion properties of these events are significantly different or not.

\subsubsection{Further detailed comparisons}
\label{sec:differences}

\begin{figure*}
  \begin{center}
    \includegraphics[angle=0,width=12.5cm]{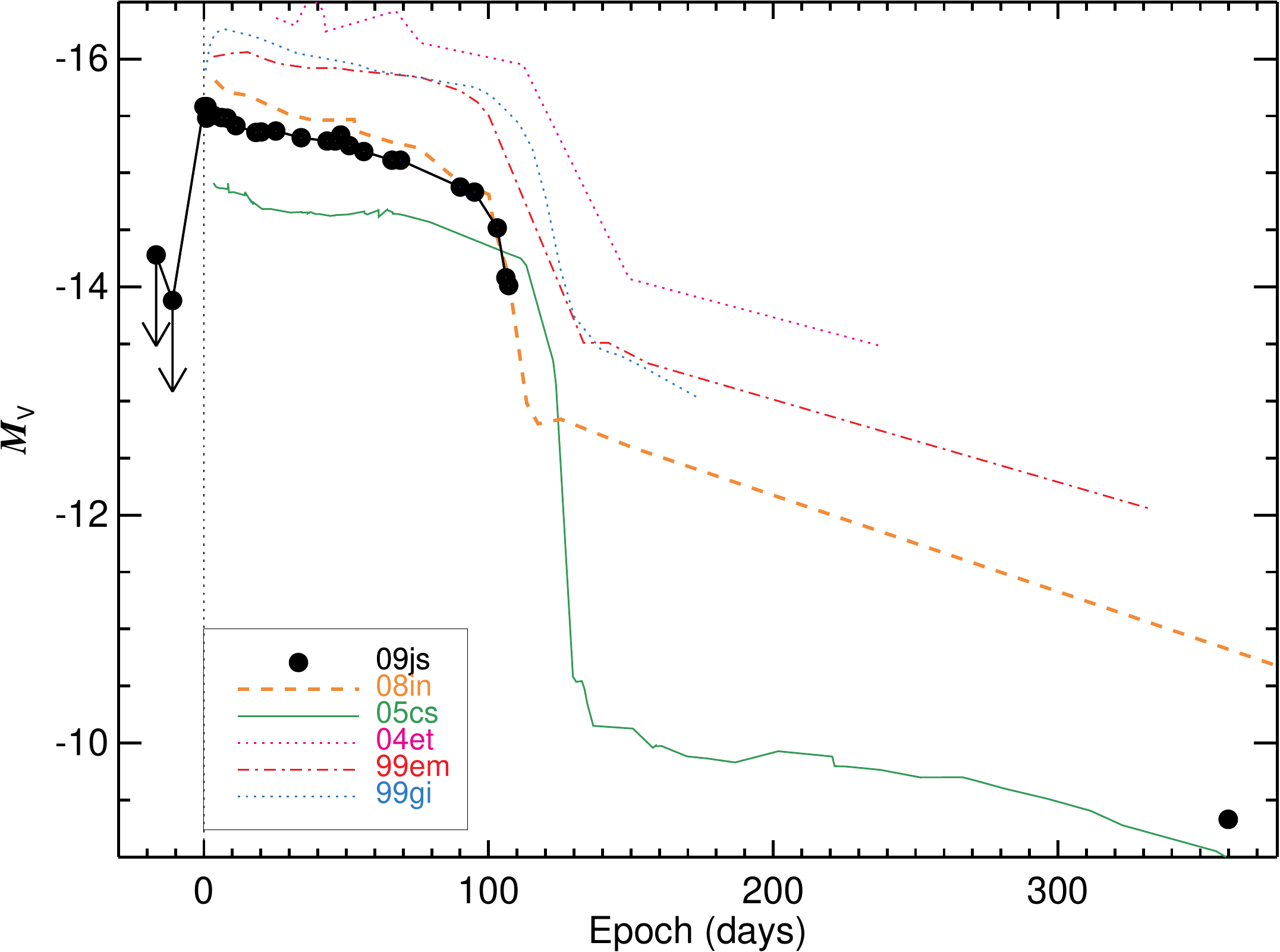}
    \caption{Absolute reddening-corrected magnitude ($M_V$) light curve of SN~2009js compared to other events. 
    \label{fig:vabs}}
  \end{center}
\end{figure*}

1. The plateau luminosity of SN~2009js is higher than that of SN~2005cs by a factor of about 2, and more closely matched to SN~2008in, as mentioned in the previous section. During transition between plateau and nebular phase, however, SN~2009js fades and falls below the luminosity of SN~2005cs. During the late nebular phase, the luminosity of SN~2009js lies in between that of SN~2005cs and SN~2008in. This is readily apparent from the quasi-bolometric luminosity light curve and the absolute $V$ band light curve in Figs.~\ref{fig:evolution} and \ref{fig:vabs}. 

\begin{figure}
  \begin{center}
  \includegraphics[angle=0,width=8.5cm]{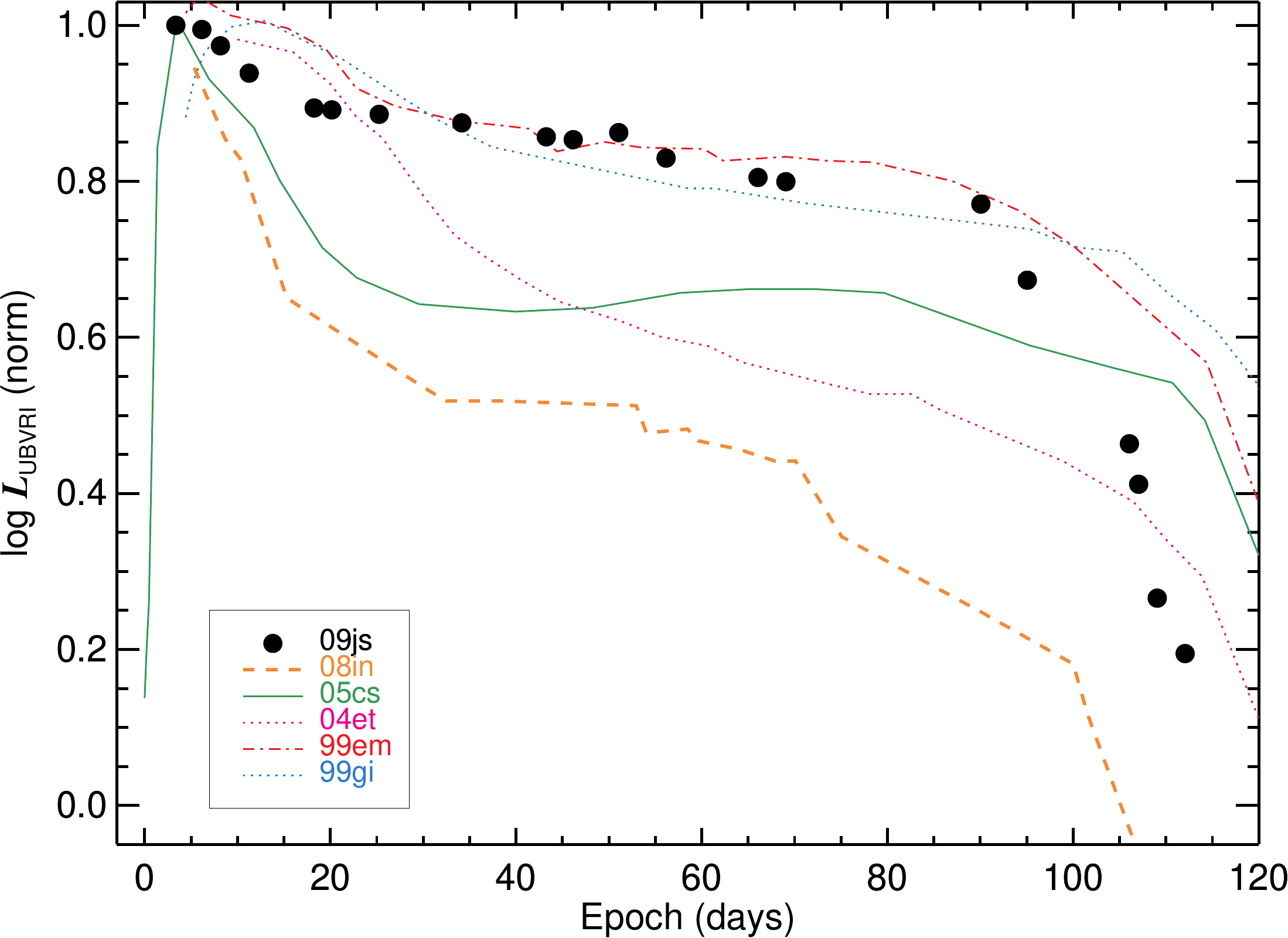}
    \caption{Evolution in quasi-bolometric power (integrated under the Planck function between 3500\AA\ and 1~\micron) after correction for local and Galactic reddening. 
    \label{fig:evolutionnorm}}
  \end{center}
\end{figure}

2. The plateau phase is somewhat shorter than that of SN~2005cs, which is why the luminosity of SN~2009js drops below that of SN~2005cs during the transition phase. In Eq. 2, we have measured $t_{\rm PT}$(2009js)=111 days. We carried out an identical fit for the light curve of SN~2005cs and found a significantly larger $t_{\rm PT}$(2005cs)=130.8\p0.2 days. The difference of about 20 days is larger than the uncertainties on the explosion epoch (\S~\ref{sec:explosionepoch}). A comparison with other SNe yields a similar result, a fact that is best visible in Fig.~\ref{fig:evolutionnorm}. This figure shows the evolution of $L_{\rm UBVRI}$ values for various SNe as in Fig.~\ref{fig:evolution}, with the only change being that all sources are shifted vertically (normalized) in order to approximately match the early-time peak of SN~2009js. Apparently, the drop in \lbol\ for SN~2009js occurs from a few days before, to up to two weeks before, that in most other sources: SNe~2005cs, 2004et, 1999em and 1999gi. However, the plateau length does appear to match that of SN~2008in, which has also been measured to be short ($\sim$98 days according to \citealt{roy11}; also apparent in Fig.~\ref{fig:fulllc}).

3. \citet{pastorello06} found unusually blue $U$--$B$ colors for SN~2005cs, corresponding to very high continuum temperatures $T$ of up to 3$\times$10$^4$ K at early times, which cooled to $T$$\sim$7500 K at 35 days. According to our multi-epoch blackbody fits (\S~\ref{sec:evolution}), $T$(2009js) is significantly lower over the entire period probed (Fig.~\ref{fig:evolution}). Fig.~\ref{fig:temperature} shows 
a reasonable match to the temperature values found for SN~1999br and SN~2008in. We do not have $U$ band monitoring data for SN~2009js, so one may question whether we miss a significant fraction of the flux and underestimate the photospheric temperature. However, our multi-band optical photometry appears to satisfactorily describe the Wien tail of the optical data (see, e.g., the decrease towards shorter wavelengths in Fig.~\ref{fig:epochs}). So we are unlikely to be missing any significant component in the $U$ band, and $T$(2009js) is then significantly lower than that of SN~2005cs. This conclusion is robust to uncertainties in host galaxy reddening, which is estimated to be low (\S~\ref{sec:reddening}, \S~\ref{sec:optspec}).
 
4. Fig.~\ref{fig:evolutionnorm} shows that SN~2005cs underwent a larger luminosity decrement between peak and plateau, as compared to SN~2009js. 
In order to quantify this, we define a new parameter $\Delta$log${\cal L}$ as the difference between the peak luminosity and that measured at 50 days post-peak in log luminosity units:

\begin{equation}
  \Delta {\rm log} {\cal L} = {\rm log}L_{(t_{\rm peak})} - {\rm log}L_{(t_{\rm peak}+50{\rm d})}
  \label{eq:deltal}
\end{equation}

We have $\Delta$log${\cal L}$(2009js)=0.16 dex and $\Delta$log${\cal L}$(2005cs)=0.36 dex, when measuring the luminosity in terms of $L_{\rm UBVRI}$. The match is even worse when compared to SN~2008in, for which $\Delta$log${\cal L}$(2008in)=0.47 dex. This indicates a comparatively lower amount of radiative cooling for SN~2009js during adiabatic expansion following shock breakout. In this respect, SN~2009js is closer matched to SN~1999em ($\Delta$log${\cal L}$=0.19) and SN~1999gi ($\Delta$log${\cal L}$=0.22). 

\subsection{Nebular phase}
\label{sec:nebular}

The NTT data reveal the late phase evolution of SN~2009js. The $VRI$ light curves in terms of observed flux (Fig.~\ref{fig:fulllc}) show that the relative decline from mid-plateau to the nebular phase is very similar to that of SN~2005cs in all three available filters. Although we do not probe the beginning of the nebular tail in our Kanata monitoring, the comparison with SN~2005cs implies that the $I$ band is likely to have covered the bulk of transition phase after the plateau. This allowed us to measure the slope of flux decline in the nebular phase in \S~\ref{sec:optlc} as $p_0$=0.009 mag day$^{-1}$, a value which lies within the range of values found for other sources in the compilation of \citet{olivares10}. 

The nebular phase is the result of energy release during radioactive decay of $^{56}$Ni. Thus, the luminosity in this phase is a good tracer of the amount of $^{56}$Ni produced. The fitted blackbody temperature in this phase is 3530\p1000 K. In terms of (quasi)bolometric optical power, $L_{\rm UBVRI}$(2009js)=2.08(\p0.59)$\times$10$^{39}$ erg s$^{-1}$. The luminosity uncertainty was determined by Monte Carlo sampling of fluxes drawn from a normal distribution with standard deviation equal to the EFOSC photometric errors in each filter, followed by integration of the fitted Planck curves to give an ensemble of randomized luminosities. This luminosity is 2.4(\p0.7)$\times$$L_{\rm UBVRI}$(2005cs) at an epoch of +360 days, and about 0.4(\p0.1)$\times$$L_{\rm UBVRI}$(2008in); in other words, SN~2009js lies between the other two SNe. 
The $^{56}$Ni mass produced in SN~2008in according to \citet{roy11} was 0.015~\Msun. Scaling this by luminosity implies $M_{\rm Ni}$(2009js)$\approx$ 0.006(\p0.002)~\Msun. Extrapolating instead from the $^{56}$Ni mass synthesized in SN~2005cs ($\sim$0.003~\Msun, according to \citealt{pastorello09}) produces a very close result, i.e. $M_{\rm Ni}$(2009js)$\approx$0.007(\p0.002)~\Msun. 
Assuming a 50\%\ higher bolometric correction (see \S~\ref{sec:evolution}), this may increase to $M_{\rm Ni}$(2009js) $\approx$0.011(\p0.003)~\Msun. 

An important caveat regarding measurement of the luminosity is that a Planck function may be a poor approximation to the source spectrum during the nebular phase. This has been investigated by \citet{hamuy01} who suggests that although such an approximation is worse in the nebular phase, the Planck function still gives comparable results to direct integration over the full optical--infrared regime, to within 15\%\ accuracy. Using the bolometric correction of 0.26 mag from the $V$-band found by \citet{hamuy01} for SN~1987A and SN~1999em, we compute a nebular phase total bolometric luminosity (\lbol) of 3.6(\p2.7)$\times$10$^{39}$ erg s$^{-1}$ which is only 1.7$\times$ larger than (and consistent within the reddening-corrected EFOSC $V$-band flux uncertainty with) the value of $L_{\rm UBVRI}$ used above and plotted in Fig.~\ref{fig:evolution}. 
If we instead use the $R$ band absolute magnitude in order to avoid uncertainties related to bolometric corrections, then we have $M_{\rm R}$(2009js)$\approx$$M_{\rm R}$(2005cs)--0.20 (with a 40\%\ reddening-corrected flux uncertainty) at an epoch of +360 days. Scaling $M_{\rm Ni}$ then implies $M_{\rm Ni}$(2009js)$\approx$0.004(\p0.0016)~\Msun.

In summary, the above range of $M_{\rm Ni}$ is much smaller than synthesized in normal SNe II (for which $M_{\rm Ni}$ is generally greater than 0.05~\Msun), supporting the association with the subluminous class of events.

\subsection{Infrared detections: photosphere or dust emission?}
\label{sec:irorigin}

The \spitzer\ detections of SN~2009js at an epoch of 2 days post-discovery, or 7.5\p5.5 days post-explosion (\S~\ref{sec:results_irphot}) represent some of the earliest mid-infrared (MIR) SN detections, especially beyond 4\micron. Only SN~1987A has comparably-early mid-infrared follow-up in the shorter-wavelength $L$ band \citep{menzies87}. Subluminous SNe have not been studied in detail in the mid-infrared so far.\footnote{Archival \spitzer\ data of SN~2005cs was studied by \citet{szalai12}, but no reliable detection was found.} The fact that mid-infrared detections of SN~2009js with both \spitzer\ and \wise\ were serendipitous (i.e. not part of any SN survey) emphasizes the importance of examining archival data and all-sky surveys in detail. 

The question of dust production in SNe is a very active topic of research. Dust has been detected in several SNe using mid- and far-infrared observations \citep[e.g. ][]{wooden93, kotak09, meikle11, matsuura11, fox11, szalai12}. The pace of dust production in core-collapse SNe is uncertain, but is generally regarded to have a characteristic timescale of several months to one year or more \citep[e.g. ][]{nozawa03}. Early-time MIR excesses have been detected in some cases (e.g. in SN~1987A by \citealt{wooden93}, and in SN~2004dj by \citealt{meikle11}). In SN~1987A, the excess is attributed to heating of circumstellar dust, while in SN~2004dj, it is instead associated with echos from newly formed dust within a cool dense shell. 

With only two MIR bands and the absence of any near-infrared monitoring, it is difficult to assess the contribution of dust to the spectral energy distributions (SEDs) of SN~2009js. However, we note that the spectral slopes between the two available mid-infrared bands are flatter than those expected from a blackbody fitted to the broadband (optical+MIR) data. This is easily visible in Fig.~\ref{fig:epochs} and is found to be true for both Kanata epochs +3.4 days and +107.1 days when MIR detections are available. In the former, the slope is only marginally flatter than the best-fitting Planck function within the photometric uncertainties. In the latter, on the other hand, the 4.6 \micron\ \wise\ W2 excess appears to be significantly larger. This may be consistent with the expectation of increased dust production on timescales of months. 

\begin{figure}[h]
  \begin{center}
  \includegraphics[width=8.5cm,angle=0]{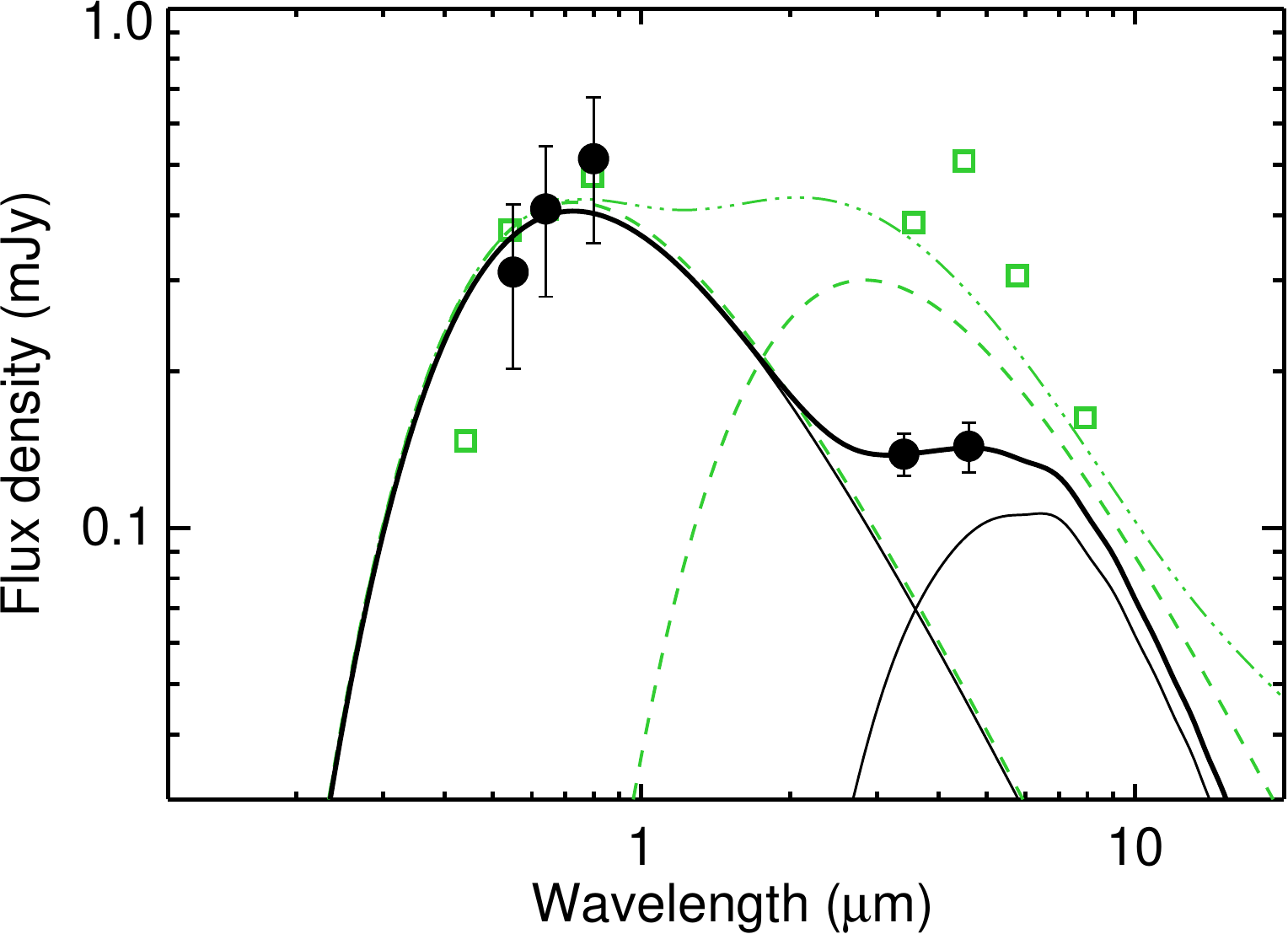}
    \caption{Comparison of the optical and MIR SEDs of SN~2009js with SN~2004dj. Two-blackbody fit to the optical and mid-IR data of SN~2009js is denoted by the black continuous curves and filled circles at an epoch of $\sim$107 days (Table~\ref{tab:kanata}). The photospheric temperatures of the two model sources have been matched. For a Carbon dust model, the mass of the dust fitting the mid-IR data is $\sim$3$\times$10$^{-5}$~\Msun. In light color and unfilled green squares, the SED and triple blackbody fit for SN~2004dj at an epoch of 106 days are shown \citep{meikle11}. These have been normalized to match SN~2009js in the optical. 
    \label{fig:dust}}
  \end{center}
\end{figure}

Although accurate assessment of the dust contribution is difficult, it is possible to compare SN~2009js with another Type IIP SN~2004dj, in which MIR excesses attributed to freshly formed dust have been observed \citep{meikle11}. Fig.~\ref{fig:dust} compares the optical and MIR SED of these two sources at a similar epoch of 106 days. For the latter source, the data and triple blackbody fit are both taken from \citet{meikle11} and are normalized to match SN~2009js in the optical. From the fact that the MIR fluxes of SN~2009js lie well below those of SN~2004dj (normalized to the optical photospheric emission), it is immediately clear that dust heating in SN~2009js is much less efficient than in SN~2004dj. In order to construct a comparative model of any potential dust emission in SN~2009js, we assumed that the photospheres of the two sources appear identical at this epoch, i.e. both have a temperature $T_{\rm photosphere}$=7000 K \citep{meikle11}. Two blackbody functions were used to represent the photosphere and the dust respectively. The fitted model is shown in Fig.~\ref{fig:dust}, and yields $T_{\rm dust}$(2009js)$\approx$630 K, which is lower by a factor of about 3 than $T_{\rm dust}$(2004dj)=1800 K from \citet{meikle11}. Assuming that the emission region is optically thin and that grain composition is mainly amorphous carbon \citep{zubko96}, we find a dust mass of $M_{\rm dust}$(2009js)$\sim$3.0$\times$10$^{-5}$~\Msun.\footnote{A predominantly Silicate dust composition, or a much cooler dust component, cannot be ruled out, but these are less likely because a prominent detection in the \wise\ W3 filter (encompassing the Silicate emission feature wavelength) is predicted in these cases at a level of $\sim$1 mJy. This flux level corresponds closely to the canonical flux limit of the \wise\ all-sky survey \citep{wise}, and there is no such detection in the all-sky release.}

The integrated dust luminosity is only about 3~\%\ of the integrated photospheric luminosity in SN~2009js, whereas it is about 15~\%\ for SN~2004dj (see Table~3 of \citealt{meikle11}). The absolute luminosity of the dust component for SN~2004dj is also higher, at about 2.5 times the value that we find for SN~2009js, even though the luminosity of the photosphere of SN~2004dj is about half that of SN~2009js at this epoch. \citet{meikle11} attributed the MIR excess to freshly-formed dust within a cool dense shell. If the same were to be true in the case of SN~2009js and if both events had similar progenitors, this would imply much less dust production in SN~2009js. But the present data cannot distinguish freshly-formed dust from a scenario where the infrared excess originates as an echo from pre-existing dust. This latter possibility may be more likely, as various statistical studies of SNe dust have indicated \citep{fox11, szalai12}. MIR detections at epochs of about 100 days may be too early for the infrared excess to be interpreted as thermal radiation from dust formed in the ejecta (in Type IIP SNe, dust generally condenses after about 300 days; e.g., \citealt{nozawa03}). 

One other possibility to consider is that the 3.4~\micron\ emission is photospheric and not dust-related (as suggested in the final panel of Fig.~\ref{fig:epochs}), whereas the 4.6~\micron\ excess is the result of fundamental CO molecular line emission. Such molecular emission is thought to be an important precursor signal to the imminent onset of dust condensation in SNe ejecta, and has been discussed in the context of several SNe (e.g., \citealt{kotak06, meikle11}).

To summarize, the comparison with SN~2004dj suggests evidence of heated dust emission in SN~2009js at an epoch of about 107 days. Alternatively, the MIR excess above the photosphere may be a sign of dust-precursor molecular emission. Both of these scenarios are interesting, and it is important to extend studies of dust formation to the subluminous SNe regime with further detailed follow-up work.

\subsection{The nature of the progenitor star and the SN explosion}
\label{sec:progenitor}

The nature of the progenitor of SN~2009js is presently unknown. The fact that the source is 2005cs--like may suggest a similar progenitor, which is constrained to be a moderate-mass (8--15 \Msun) star \citep{pastorello09}. Progenitors of subluminous events may generally lie at the lower mass range of stars that produce core-collapse events (e.g. \citealt{fraser11}). The mass constraint on the intermediate luminosity source
SN~2008in is obtained only from analytic modeling of 
the light curve and spectra: an upper-limit of 20~\Msun\ \citep{roy11}. 

We have \spitzer\ pre-explosion images for SN~2009js, but these do not provide stringent constraints on the progenitor. We instead use established analytical relations in order to approximately estimate the explosion energy ($E$), ejected mass ($M_{\rm ej}$) and the pre-SN stellar radius ($R_{\rm pre-SN}$). For this purpose, Eqs.~1--3 of \citet[][ see also \citealt{popov93}]{litvinova85} may be used, with the caveats that these relations ignore some important effects such as nickel heating and line opacities. The required observables are the plateau length ($\Delta t$) and the absolute $V$ magnitude $M_{\rm V}$ at mid-plateau, and the photospheric velocity ($u_{\rm ph}$) also at mid-plateau. For SN~2009js, we have $\Delta t$$\approx$111\p 5.5 days. This is taken to be $t_{\rm PT}$ (Eq. 2), assuming that the plateau begins after the initial rise of a few days (which we do not probe), and also includes the uncertainty on the determination of the explosion epoch. $M_{\rm V}$=--15.2\p0.3 at an epoch of 55 days. The absence of spectroscopic monitoring means that $u_{\rm ph}$ at mid-plateau is unknown. But the good spectral template match to SN~2005cs (cf. \S~\ref{sec:2005cslike}) suggests that using an identical velocity evolution is a reasonable way to proceed. Interpolating the Sc {\sc ii} velocity evolution for SN~2005cs found by \citet{pastorello09} at an epoch of about +55 days, we find $u_{\rm ph}$$\sim$1600 km s$^{-1}$ to which we assign a generous systematic uncertainty of 400 km s$^{-1}$ to account for the absence of observations for SN~2009js.\footnote{Such an uncertainty would also encompass the Fe {\sc ii} velocity observed by \citet{roy11} for SN~2008in at a similar epoch, at the 2$\sigma$ confidence level.} This yields 

\begin{eqnarray}
  E & = & 0.14\pm 0.11\ {\rm foe}\\
  M_{\rm ej} & = & 8.9\pm 4.8\ M_{\odot}\\
  R_{\rm pre-SN} & = & 415\pm 340\ R_{\odot}.
\end{eqnarray}

The ejecta mass is about 9~\Msun. Adding about 1~\Msun\ for stellar wind mass losses and another 1.4~\Msun\ for the compact remnant assumed to collapse during the SN, the initial progenitor mass is estimated to be about 11\p5~\Msun. \citet{maguire10} have shown that progenitor mass estimates based upon analytical relations are generally higher than those based upon pre-explosion images or detection limits, and some additional caution is thus required when using these values. However, a useful statement which can be made is that these estimates of $M_{\rm ej}$ and $M_{\rm progenitor}$ are consistent with those derived for SN~2005cs using various methods \citep{pastorello09, maguire10}.

The explosion energy derived above for SN~2009js is lower than that for SN~2008in (0.5 foe; \citealt{roy11}). The value of $E$ derived similarly for SN~2005cs (0.17\p0.08 foe) by \citet{maguire10}, on the other hand, is consistent with that of SN~2009js, using identical analytical estimates as above. These results provide additional support of a close association between SN~2009js and SN~2005cs.\footnote{keeping the caveat in mind that the $E$ estimate is quite sensitive to $u_{\rm ph}$, for which we do not have a direct measurement.}

\section{Summary}

Early-time photometry previously reported constrains the explosion epoch of SN~2009js to MJD 55109.94\p5.5 days (\S~\ref{sec:explosionepoch}). 

Kanata optical monitoring shows SN~2009js to have characteristics typical for a Type IIP event (\S~\ref{sec:evolution}). The characteristic plateau length (as measured by the mid-point of the transition between plateau and nebular phases) is 111 days (\S~\ref{sec:optlc}). Line-of-sight extinction is found to be \av(Galaxy)$\approx$0.95 mags (\S~\ref{sec:galreddening}) and \av(host galaxy)=0.18\p 0.38 mags (\S~\ref{sec:reddening}), i.e. the reddening local to the SN is lower than that due to our Galaxy. 

Subaru optical spectroscopy shows a very good match to the spectrum of the well-studied subluminous Type IIP SN~2005cs (\S~\ref{sec:optspec}). These facts suggest that SN~2009js was a SN~2005cs--like event (\S~\ref{sec:2005cslike}).

We are able to chart the evolution of the photospheric temperature and optical bolometric power of SN~2009js (\S~\ref{sec:evolution}). The plateau-phase luminosity of SN~2009js is higher than that of SN~2005cs, but significantly lower than normal Type IIP events such as SN~1999em (Fig.~\ref{fig:evolution}) and SN~1999gi (Fig.~\ref{fig:vabs}). Some other differences with respect to SN~2005cs are also noted (\S~\ref{sec:differences}). The initial post-peak adiabatic phase cooling is less prominent than in SN~2005cs, as quantified by a new parameter $\Delta$log${\cal L}$ measuring the difference in log luminosities at peak, and 50 days post-peak. Moreover, the plateau length of SN~2009js is significantly shorter than in SN~2005cs, and more similar to that of the intermediate luminosity event SN~2008in. The absolute mid-plateau luminosity of SN~2008in also closely matches that of SN~2009js. On the other hand, SN~2008in shows a significantly higher nebular luminosity, as well as a larger cooling dip $\Delta$log${\cal L}$. In addition, its photospheric velocity determined from spectroscopy is significantly higher than SN~2005cs and SN~2009js. These facts imply that SN~2009js has characteristics common to both subluminous and intermediate luminosity events. 

The mass of synthesized radioactive $^{56}$Ni is found to be 0.004--0.011~\Msun\ (\S~\ref{sec:nebular}). This is similar to the synthesis in other subluminous events. 

Using known analytical relations, a low explosion energy $E$=0.14\p 0.11 foe is derived. This is lower than for SN~2008in, within the caveats of use of the analytical relations. The constraints on the ejecta mass (8.9\p4.8~\Msun) and progenitor mass, as well as on the explosion energy, are fully consistent with those derived for SN~2005cs using similar methods (\S~\ref{sec:progenitor}). 

SN~2009js is the first subluminous (or intermediate luminosity) SN to be studied in the mid-infrared (MIR). \spitzer\ detected SN~2009js serendipitously at a very early epoch, and \wise\ monitored the field on three occasions to an epoch of about +470 days. A significant MIR excess above the photosphere is found at 4.6~\micron\ at an epoch of $\sim$107 days post-discovery. A tentative dust model fit implies a mass of amorphous Carbonaceous dust of about 3$\times$10$^{-5}$~\Msun, and a low efficiency of dust heating as compared to SN~2004dj (\S~\ref{sec:irorigin}). Alternatively, this may be a sign of fundamental CO molecular emission which is a precursor to dust formation.

Low luminosity events constitute only a few percent of the Type II SN population \citep[e.g. ][]{pastorello04}, and few studies exist on how these connect to the \lq normal\rq\ population. Our observations of SN~2009js thus add to the list of sources in this class with relatively well-sampled multi-wavelength light curves and provide first insight into their mid-infrared properties.

\acknowledgements
PG acknowledges a JAXA International Top Young Fellowship. TJM is supported by the Japan Society for the Promotion of Science Research Fellowship for Young Scientists $(23\cdot5929)$. This research is supported by World Premier International Research Center Initiative, MEXT, Japan. We thank the referee for their report, which helped to streamline the text flow, correct important typos and to make the discussion more robust.

This publication makes use of data products from the Wide-field Infrared Survey Explorer, which is a joint project of the University of California, Los Angeles, and the Jet Propulsion Laboratory/California Institute of Technology, and NEOWISE, which is a project of the Jet Propulsion Laboratory/California Institute of Technology. WISE and NEOWISE are funded by the National Aeronautics and Space Administration. This work is based in part on observations made with the \spitzer\ Space Telescope, obtained from the NASA/ IPAC Infrared Science Archive, both of which are operated by the Jet Propulsion Laboratory, California Institute of Technology under a contract with the National Aeronautics and Space Administration. 

This research has made use of SAOImage DS9 \citep{ds9}, developed by Smithsonian Astrophysical Observatory. This research has made use of NASA's Astrophysics Data System Bibliographic Services.

\appendix

\section{Optical photometry}

\subsection{Kanata optical imaging}

The 1.5 m Kanata telescope \citep{kanata}, part of the Higashi-Hiroshima Observatory at an altitude of 503-m above sea level, is used by our group for regular monitoring of SNe. HOWPol $BVR_{\rm C}I_{\rm C}$ imaging provides a 15\arcmin--diameter field of view at a pixel scale of 0\farcs 29. 

Monitoring started on 2009 Oct 14.8 at an epoch of +3.4 days and continued for about 4 months. A total of 25 epochs of data on different nights were obtained during this time. 

The images were
reduced according to normal procedures for CCD photometry.
We performed point-spread-function fitting photometry using the
$DAOPHOT$ package in \iraf\footnote{http://iraf.noao.edu}.  We also observed 
PG~0231$+$051 as a standard star field from the Landolt standard
star catalog \citep{landolt92} when the night was photometric. The magnitudes of
local comparison stars\footnote{These stars have the following IDs in the USNO-B1.0 survey \citep{usno}: 1084-0033855, 1084-0033919 and 1085-0027773.} and the color terms were calibrated using 
these standard stars. Photometric uncertainties include Poisson measurement error and standard deviation resulting from calibration against different comparison stars. Zeropoints are those from the canonical filter definitions of \citep{bessell98}.

\subsection{NTT late-time optical imaging}

NTT/EFOSC was used for observing SN~2009js on 2010 Oct 6.\footnote{Based on observations obtained during ESO program 184.D-1140(E), P.I. S. Benetti.} Observing conditions were cloudy with a median seeing of about 1\farcs 9 in the $V$ band. A series of five consecutive exposures (100 s each, giving a total exposure time per filter of 500 s) were taken in $VRI$ starting at UT05:53 at a median airmass of 1.48. 

Reduction was carried out using the EFOSC2 pipeline and standard procedures for analysis \citep{izzo11}. The reduced exposures were combined to a cleaned final image. Since the source is relatively faint at this late epoch, the background was carefully measured using adjacent apertures that include emission from the sky as well as the underlying host galaxy. No photometric standard star observations were obtained that night. We therefore carried out relative photometry with respect to nearby comparison stars, and then tied the absolute calibration to the Kanata data using stars common to the Kanata and NTT images. This also alleviates the issue of small differences in filter response between the $VRI$ filters of the two telescopes. 
The resultant photometry is listed in Table~\ref{tab:kanata}.

\subsection{Early filterless photometry}

In addition to the above datasets analyzed by us, we also made direct use of pre-discovery and discovery photometry reported in CBET 1969 \citep{nakano09, silverman09}. These data were obtained in filterless CCD mode and are listed at the top of Table~\ref{tab:kanata}. Although unfiltered CCDs have broad response functions, their effective central wavelengths are typically close to that of the $V$ band. \citet{riess99} have determined cross-calibration equations between unfiltered and $V$ band photometry, based upon the $B-V$ color. 
Although we do not know the SN colors at the discovery epoch, the first Kanata measurements corrected for reddening (see \S~\ref{sec:galreddening} and \ref{sec:reddening}) imply $B-V$$\approx$0.3, as expected for a hot blackbody characterizing the early-time SN spectrum. Using Eq.~3 of \citet{riess99} and the transformation constant from Table~2 therein implies that the unfiltered mag is expected to be within 0.1 mags of the $V$-band mag. We thus use the unfiltered mags directly as a proxy for the $V$ band, with a systematic uncertainty of 10\%.

\begin{table*}
\begin{center}
 \caption{Optical photometry\label{tab:kanata}}
   \begin{tabular}{lccccc}
     \hline
Epoch$^{\dag}$   & MJD   &         $B$        &         $V$        &         $R$        &        $I$       \\
               &       &       mag~~~~~~~(mJy)       &         mag~~~~~~~(mJy)        &         mag~~~~~~~(mJy)        &        mag~~~~~~~(mJy)        \\
     \hline
&&&&&\\
\multicolumn{6}{c}{\underline{\textsl{Early-time unfiltered}}}\\
--16.8   &   55098.7    &    --    &        $<$18.5$^I$    &--    &    --   \\
--11.0   &   55104.4    &    --    &        $<$18.9$^K$    &--    &    --   \\
0.0      &   55115.4    &    --    &        17.2$^K$    &--    &    --   \\
0.3      &   55115.7    &    --    &        17.2$^I$    &--    &    --   \\
0.7      &   55116.2    &    --    &        16.7$^Y$   &--    &    --   \\
1.0      &   55116.4    &    --    &        17.3$^K$    &--    &    --   \\
1.1      &   55116.5    &    --    &        17.2$^I$        &--    &    --   \\
&&&&&\\
\multicolumn{6}{c}{\underline{\textsl{Kanata}}}\\
  3.4   &   55118.8    &    17.95\p 0.05 (1.04\p0.39)    &    17.28\p 0.04 (1.23\p0.45)    &    --    &    --   \\
  6.2   &   55121.6    &    17.97\p 0.05 (1.03\p0.38)    &    17.30\p 0.02 (1.21\p0.44)    &    16.76\p 0.03 (1.37\p0.49)    &    16.49\p 0.03 (1.08\p0.38)   \\
  8.2   &   55123.6    &    18.07\p 0.10 (0.93\p0.35)    &    17.30\p 0.10 (1.20\p0.45)    &    16.79\p 0.04 (1.34\p0.48)    &    16.52\p 0.03 (1.06\p0.37)   \\
 10.2   &   55125.6    &    18.16\p 0.04 (0.86\p0.32)    &    --    &    --    &    --   \\
 11.3   &   55126.7    &    18.25\p 0.05 (0.79\p0.29)    &    17.37\p 0.04 (1.13\p0.41)    &    16.80\p 0.04 (1.32\p0.47)    &    16.51\p 0.03 (1.07\p0.38)   \\
 18.3   &   55133.7    &    18.56\p 0.12 (0.59\p0.23)    &    17.43\p 0.06 (1.07\p0.39)    &    16.83\p 0.04 (1.28\p0.46)    &    16.46\p 0.07 (1.11\p0.40)   \\
 20.2   &   55135.6    &    18.60\p 0.30 (0.57\p0.26)    &    17.42\p 0.13 (1.08\p0.41)    &    16.80\p 0.09 (1.32\p0.48)    &    16.48\p 0.06 (1.09\p0.39)   \\
 24.2   &   55139.6    &    --    &    --    &    16.87\p 0.05 (1.24\p0.45)    &    16.43\p 0.06 (1.15\p0.41)   \\
 25.3   &   55140.7    &    18.68\p 0.17 (0.53\p0.21)    &    17.41\p 0.05 (1.08\p0.39)    &    16.83\p 0.04 (1.29\p0.46)    &    16.43\p 0.04 (1.15\p0.41)   \\
 34.2   &   55149.6    &    18.83\p 0.10 (0.46\p0.18)    &    17.47\p 0.06 (1.03\p0.37)    &    16.83\p 0.04 (1.28\p0.46)    &    16.35\p 0.04 (1.23\p0.44)   \\
 43.3   &   55158.7    &    19.02\p 0.11 (0.39\p0.15)    &    17.50\p 0.04 (1.00\p0.36)    &    16.85\p 0.04 (1.26\p0.45)    &    16.33\p 0.03 (1.26\p0.45)   \\
 46.2   &   55161.6    &    19.06\p 0.11 (0.38\p0.14)    &    17.50\p 0.04 (1.00\p0.36)    &    16.85\p 0.06 (1.27\p0.46)    &    16.33\p 0.03 (1.25\p0.44)   \\
 48.1   &   55163.5    &    --    &    17.45\p 0.17 (1.05\p0.41)    &    16.85\p 0.13 (1.27\p0.48)    &    16.35\p 0.08 (1.24\p0.45)   \\
 51.1   &   55166.5    &    18.99\p 0.32 (0.40\p0.19)    &    17.54\p 0.09 (0.96\p0.36)    &    16.85\p 0.05 (1.27\p0.46)    &    16.33\p 0.04 (1.26\p0.45)   \\
 56.2   &   55171.6    &    19.29\p 0.14 (0.30\p0.12)    &    17.59\p 0.05 (0.92\p0.33)    &    16.87\p 0.05 (1.24\p0.45)    &    16.32\p 0.04 (1.27\p0.45)   \\
 66.1   &   55181.5    &    19.36\p 0.21 (0.28\p0.12)    &    17.67\p 0.04 (0.86\p0.31)    &    16.95\p 0.05 (1.16\p0.42)    &    16.38\p 0.05 (1.20\p0.43)   \\
 69.1   &   55184.5    &    19.49\p 0.15 (0.25\p0.10)    &    17.67\p 0.05 (0.86\p0.31)    &    16.94\p 0.03 (1.16\p0.42)    &    16.36\p 0.03 (1.23\p0.43)   \\
 90.1   &   55205.5    &    --    &    17.90\p 0.09 (0.69\p0.25)    &    17.17\p 0.07 (0.94\p0.34)    &    16.62\p 0.04 (0.96\p0.34)   \\
 95.1   &   55210.5    &    19.92\p 0.21 (0.17\p0.07)    &    17.95\p 0.08 (0.66\p0.24)    &    17.28\p 0.06 (0.85\p0.31)    &    16.64\p 0.05 (0.95\p0.34)   \\
103.1   &   55218.5    &    --    &    18.26\p 0.20 (0.50\p0.20)    &    --    &    --   \\
106.1   &   55221.5    &    --    &    18.70\p 0.27 (0.33\p0.14)    &    17.93\p 0.15 (0.47\p0.18)    &    17.29\p 0.10 (0.52\p0.19)   \\
107.1   &   55222.5    &    --    &    18.77\p 0.16 (0.31\p0.12)    &    18.07\p 0.09 (0.41\p0.15)    &    17.30\p 0.05 (0.51\p0.18)   \\
109.1   &   55224.5    &    --    &    --    &    18.47\p 0.16 (0.28\p0.11)    &    17.62\p 0.09 (0.38\p0.14)   \\
112.1   &   55227.5    &    --    &    --    &    18.66\p 0.21 (0.24\p0.10)    &    17.88\p 0.17 (0.30\p0.12)   \\
117.1   &   55232.5    &    --    &    --    &    18.87\p 0.13 (0.20\p0.07)    &    18.59\p 0.20 (0.16\p0.06)   \\
&&&&&\\
\multicolumn{6}{c}{\underline{\textsl{EFOSC}}}\\
359.8   &   55475.3    &    --    &    23.45\p 0.63 (0.004\p0.003)    &    21.95\p 0.30 (0.010\p0.004)    &    21.10\p 0.10 (0.014\p0.003)   \\
&&&&&\\
     \hline
   \end{tabular}
~\\
$^\dag$Denotes days post-discovery (2009 Oct 11.44).\\
For the unfiltered early-time photometry, the observers are Itagaki (denoted by superscript $I$), KAIT (superscript $K$) and Yusa (superscript $Y$). The measurement by Yusa is not included in our analysis because it is brighter by 0.5 mags as compared to all other photometry obtained both before and after. No uncertainties are reported and it is impossible to judge its veracity.
\\
Listed magnitudes and magnitude limits are the observed ones in the $B$, $V$, $R$ and $I$ bands, whereas fluxes have been corrected for Galactic and host reddening. Errors in flux (1$\sigma$) include corresponding systematic dereddening uncertainties.\\
 \end{center}
\end{table*}

\section{Optical spectroscopy}

SN 2009js was observed on UT 2009-10-27 (MJD = 55131.6),
about 16 days post-discovery with the Subaru telescope
equipped with Faint Object Camera and Spectrograph
(FOCAS; \citealt{focas}).
The blue (3800-7000 \AA) and red (6000-10000 \AA) parts 
of the spectrum were taken separately.
The blue part was taken 
with the 300B grism without an order-cut filter,
while the red part was taken with the 300R grism with the O58 filter.
These configurations give a wavelength resolution of 10 \AA \
(500 $\rm km\ s^{-1}$ near the center of the spectrum, 
$R = \lambda/\Delta \lambda \simeq 600$).
For both settings, we used a center slit of 0\farcs 8 width.
The target was integrated for 600 s in both settings. 
The typical seeing during the observations was 0\farcs 8 in the $V$ band,
with median airmass of about 1.65. Reduction was carried out in {\sc iraf}. 
Wavelength calibration was performed using the Th-Ar arc lamp.
Flux calibration was carried out using spectrophotometric star observations
on the same night.
Finally, the blue and red part of the spectrum were combined.

\section{Infrared photometry}

\subsection{\spitzer\ mid-infrared imaging}
Archival MIR imaging data of NGC~918 are available from the \spitzer\ archive in both bands IRAC1 and IRAC2. These were taken in late 2009, a few months into the warm mission phase. The instrument point response functions are complex, but imaging quality is better than 2\arcsec\ in terms of equivalent Gaussian full-width at half-maximum measurements\footnote{http://irsa.ipac.caltech.edu/data/SPITZER/docs/irac/iracinstrumenthandbook/5}. Level 2 post-BCD IRAC mapping mosaics from the post-cryogenic phase of the mission, processed by standard pipeline software version S18.12.0, were downloaded and directly used for the present analysis. 

\spitzer\ happened to observe the galaxy at epochs of about --35 days and +2 days. For each of the IRAC1 and IRAC2 bands, the pre-explosion epoch image was subtracted from the post-discovery one to remove background structured emission from the host galaxy. Source photometry was carried out in this subtracted image using a large aperture to encompass the total flux. Conversion from image units to final fluxes was done using standard procedures and conversion factors\footnote{http://irsa.ipac.caltech.edu/data/SPITZER/docs/irac/iracinstrumenthandbook/IRAC\_Instrument\_Handbook.pdf}. As a cross-check, we also carried out detection of all sources within a few arcmin of the host galaxy nucleus using the SEXtractor package \citep{sextractor} and computed a zeropoint for each image based upon cross-calibration with objects common to the \wise\ all-sky catalog for which all magnitudes are publicly-available. The bandpass difference between IRAC1 and \wise\ W1 is small and was ignored \citep{jarrett11}. Similarly, IRAC2 and \wise\ W2 bandpass differences were ignored. We found our cross-check to be consistent with the direct \spitzer\ photometry. 

The log of observations is listed in Table~\ref{tab:obslog}, images are presented in Fig.~\ref{fig:spitzer} and photometry tabulated in Table~\ref{tab:irphot}. Non-detection upper-limits in the pre-explosion images were determined by the addition of artificial point sources using the {\tt addstar} routine in \iraf. This procedure involved generation of a point spread function (PSF) from nearby stars of known magnitude, followed by an iterative procedure of injecting scaled versions of the PSF into the coadded images at the position of the SN and running automatic source detection using the \sextractor\ package \citep{sextractor} to determine the faintest reliable artificial source detection for which we assumed a 2$\sigma$ detection threshold.

\subsection{\wise\ mid-infrared imaging}
\label{sec:wise}
The {\em Wide-field Infrared Survey Explorer} (\wise) satellite \citep[][]{wise} carried out sky surveys in four bands (W1--W4 centered on wavelengths of $\approx$3.4, 4.6, 12 and 22 \micron, respectively), at an effective angular resolution of $\approx$6\arcsec\ in W1--W3, and 12\arcsec\ in W4. Based upon the cryogen status, the publicly-released data products are divided into 1) an all-sky release, 2) a 3-band cryo release, and 3) a NEOWISE post-cryo release. Combined, these cover the period of 2010 Jan 7 to 2011 Feb 1\footnote{http://wise2.ipac.caltech.edu/docs/release/allsky/expsup/}. The sensitivity in the two short wavelength bands (W1 and W2) did not change dramatically with time, whereas the nominal sensitivities in W3 and W4 is degraded in the latter two releases. As of mid-2012, Atlas images as well as catalogs of measured positions, fluxes and other parameters of pipeline-detected sources are available for the all-sky release. These are the product of coaddition of several frames and have undergone multiple levels of calibration; they are referred to as Level 3 (L3) products. For the post-cryo data, only a preliminary release is available, including individual single-pass calibrated (level 1; L1b) data. SN~2009js is not included in the catalog of pipeline-detected sources of any release.\footnote{In fact, it turns out that SN~2009js was detected by the standard \wise\ pipeline on a single (L1b) exposure observed on MJD~55222.707859. The unique detection identifier {\tt source\_id}=01305a135-002325. The cataloged mags are $W1$=14.632\p0.086 and $W2$=14.038\p0.198. However, we do not believe these to be reliable because the detections are reported only for the L1b catalog, but do not appear in the final (L3) catalog obtained after image combination. Given the short integrations of the single exposures (7.7 s), this does not appear to be consistent. Instead, we believe this detection to be that of clumpy region of the host galaxy spiral arm itself. This is consistent with the fact that the pipeline-reported goodness-of-fit reduced $\chi^2$ values for the profile-fit photometry of the single detection are bad ($\approx$4.3 and 2.4 for $W1$ and $W2$, respectively).}

Available \wise\ images for the field of NGC~918 cover three main epochs, starting from about 107 days after the discovery of SN~2009js. Subsequent epochs are separated by approximately six months. These are detailed in Table~\ref{tab:obslog}. No pre-explosion \wise\ image for subtraction is available, so photometry at the position of the SN was carried out directly on Atlas and post-cryo images covering the SN position. 

For the fully-calibrated Atlas products, we used procedures recommended by the \wise\ team.\footnote{http://wise2.ipac.caltech.edu/docs/release/allsky/expsup/sec2\_3f.html} Small aperture photometry was carried out at the SN position as well as on bright nearby field stars with cataloged magnitudes used for relative calibration. The background was carefully selected using multiple adjacent apertures in order to avoid other point sources while encompassing local host galaxy emission. Aperture corrections, as well as correlated and confusion noise estimation for uncertainty measurements were included according to the instructions in the \wise\ explanatory supplement\footnote{Ibid.}. Magnitudes were converted to fluxes using standard zeropoints \citep{wise}. 

The publicly-released Atlas (L3) image is made by coaddition of multiple exposures obtained in 2010 Jan, plus some exposures taken in 2010 Aug. We also produced our own coadded images for each of the three main epochs, starting from the L1b frames. First, the L1b data were sieved by removal of low quality frames affected by elevated background due to Moon proximity and other artifacts\footnote{http://wise2.ipac.caltech.edu/docs/release/allsky/expsup/sec1\_4d.html\#lowqual\_images}. The {\tt swarp} package \citep{terapix_swarp} was then used for background subtracting, resampling, shifting and median combining the good frames. Upper limits were determined identically as with \spitzer. 

Final images from the various WISE epochs are shown in Fig.~\ref{fig:wise} and photometry is listed in Table~\ref{tab:irphot}.

\begin{table*}[h]
\begin{center}
 \caption{Infrared Observation log\label{tab:obslog}}
   \begin{tabular}{lcccr}
     \hline
     Date (MJD)               &Telescope/Instrument/Survey & Obs. identifier   &  Bands                                      & Exposure time \\
      UT                      &                            &                   &                                             &      s        \\
                               \hline
     2009 Sep 06.7 (55080.7)  &\spitzer/IRAC/S4G           & AOR=30590976      &   3.6~\micron\ (IRAC1), 4.5~\micron\ (IRAC2)  &      23.6, 26.8\\
     2009 Oct 13.6 (55117.6)  &\spitzer/IRAC/S4G           & AOR=30606848      &   3.6~\micron\ (IRAC1), 4.5~\micron\ (IRAC2)  &      23.6, 26.8\\
     2010 Jan 26.1 (55222.1)$^\dag$  &\wise/All-sky release       &                   &   3.4~\micron\ (W1), 4.6~\micron\ (W2)        &      100.1$^\dag$ \\
     2010 Aug 02.8 (55410.8)$^\dag$  &\wise/All-sky release       &                   &   3.4~\micron\ (W1), 4.6~\micron\ (W2)        &      177.1$^\dag$ \\
     2011 Jan 25.7 (55586.7)$^\dag$  &\wise/Post-cryo release     &                   &   3.4~\micron\ (W1), 4.6~\micron\ (W2)        &      84.7$^\dag$ \\
     \hline
   \end{tabular}
~\\
$^\dag$The date quoted for the \wise\ images is the approximate median date of multiple exposures observed around that epoch. The Exposure time is simply the frame time (7.7 s) multiplied by the number of L1b frames selected for coaddition. 
 \end{center}
\end{table*}

\begin{table*}[h]
  \begin{center}
    \caption{Infrared photometry of SN~2009js\label{tab:irphot}}
    \begin{tabular}{lcccr}
      \hline
      Days$^\dag$ &     MJD   &   Telescope    &  Short MIR$^a$    & Longer MIR$^b$ \\
                 &           &                &  $\mu$Jy     &   $\mu$Jy  \\
      \hline              
      --34.8     &  55080.7 &  \spitzer      &     $<$9      &    $<$18      \\
      +2.2       &  55117.6 &  \spitzer      &    226\p13     & 211\p11 \\
      +106.7     &  55222.1 &  \wise         &     129\p12    & 136\p15      \\ 
      +295.4     &  55410.8 &  \wise          &    $<$60     &  $<$80   \\
      +471.3     &  55586.7 &  \wise          &    $<$90     &  $<$100   \\
      \hline
    \end{tabular}
    ~\par
    $^\dag$Denotes days post-discovery (2009 Oct 11.44).\\ 
    $^a$Refers to the \spitzer\ 3.6~\micron\ or \wise\ 3.4~\micron\ band.\\
    $^b$Refers to the \spitzer\ 4.5~\micron\ or \wise\ 4.6~\micron\ band.\\
    Quoted fluxes are directly-observed values, uncorrected for extinction.
  \end{center}
\end{table*}


\bibliographystyle{apj}

\label{lastpage}
\end{document}